\pdfoutput=1

\documentclass[12pt]{article}
\usepackage{jheppub}

\usepackage{amsmath}
\usepackage{amssymb}
\usepackage{graphicx,color,slashed,braket}
\usepackage{bbm}
\usepackage{ifpdf}
\usepackage{comment}
\usepackage{array}
\usepackage{float}

\newcommand{\be}{\begin{equation}}
\newcommand{\ee}{\end{equation}}
\newcommand{\ba}{\begin{eqnarray}}
\newcommand{\ea}{\end{eqnarray}}

\newcommand{\lp}{\left(}
\newcommand{\rp}{\right)}

\newcommand{\w}{\wedge}

\title{Scale separation from O-planes}

\author{George Tringas}
\author{and Timm Wrase}

\affiliation{Department of Physics, Lehigh University, 16 Memorial Drive East, Bethlehem, PA 18018, USA\\}

\emailAdd{georgios.tringas@lehigh.edu}
\emailAdd{timm.wrase@lehigh.edu}

\notoc

\abstract{
\noindent
Orientifold planes play a crucial role in flux compactifications of string theory, and we demonstrate their deep connection to achieving scale-separated solutions.
Specifically, we show that when an orientifold plane contributes at leading order to the non-zero value of the scalar potential, then either the weak coupling limit or the large volume limit implies scale separation, meaning that the Kaluza-Klein tower mass decouples from the inverse length scale of the lower-dimensional theory.
Notably, in the supergravity limit such solutions are inherently scale-separated. This result is independent of the spacetime dimension and the dimensionality of the O$p$-plane as long as $p<7$.
Similarly, we show, extending previous results, that parametric scale separation is not possible for isotropic compactifications with a leading curvature term that generically arise in the AdS/CFT context.
We classify all possible flux compactification setups in both type IIA and type IIB string theory for O$p$-planes with $2\leq p\leq 6$ and present their universal features.
While the parametrically controlled scale-separated solutions are all AdS, we also find setups that allow for dS vacua. We prove that flux quantization prevents these dS vacua in isotropic compactifications from arising in a regime of parametric control.
}

\begin{document}

\maketitle

\newpage

\section{Introduction}\label{sec:introduction}
Given the many spatial dimensions in superstring theory, a primary aim of string phenomenology is to find compactifications where the extra dimensions are much smaller than the external spacetime. In practice, this requires separating the masses of the Kaluza-Klein towers from the inverse length scale of the effective lower-dimensional theory, a phenomenon that is known as scale separation. Over the last two decades, numerous examples have been proposed within the supergravity limit of string theory~\cite{VanRiet:2023pnx, McAllister:2023vgy}.

In recent years, scale-separated vacua have received renewed attention, especially following the 2019 conjecture that compactifications to AdS space cannot be scale-separated \cite{Lust:2019zwm}. The challenge of finding trustworthy dS vacua in string theory has further spurred this line of research and its status was recently reviewed in \cite{Coudarchet:2023mfs}.

Due to ongoing debates about existing scale-separated solutions in supergravity flux compactifications, it is important to identify their minimal ingredients. In this paper, we show that if a compactification includes an orientifold contributing at leading order to the value of the scalar potential, then scale separation naturally follows in the large volume or weak coupling limits. Concretely, let the overall volume of the internal $(10-D)$-dimensional space in string frame be $vol_{10-D} = \rho^{\frac{10-D}{2}}$. Assuming that the inverse length scale squared ($1/L_{\langle V\rangle}^2 \sim \langle V\rangle$) is of the same order as the O$p$-plane contribution $V_{\text{O}p}$, we obtain a ratio independent of the spacetime dimension $D$ 
\begin{equation}\label{eq:scaleseparation} 
\frac{\langle V \rangle }{m^2_{\text{KK}}} \sim \frac{V_{\text{O}p}}{m^2_{\text{KK}}} \sim \rho^{\frac{p-7}{2}} e^{\phi} \,. 
\end{equation}
Here, $e^\phi=g_s$ denotes the string coupling, and $m_{\text{KK}}$ is the mass scale of the Kaluza-Klein tower associated with the overall internal volume. In certain cases, additional KK-towers, warped regions, etc. might have different scalings, so $m_{\text{KK}}$ above does not always denote the scale at which the EFT breaks down. However, the overall KK-scale $m_{\text{KK}}$ above is usually the scale at which our lower dimensional EFT breaks down and always needs to be decoupled from the inverse length scale associate to our EFT. This is achieved through equation \eqref{eq:scaleseparation} in the limits $g_s \ll 1$ or $\rho \gg 1$ for $p<7$. Since essentially all trustworthy solutions reside in the large volume and weak coupling regime, scale separation follows automatically whenever an O$p$-plane contributes as a leading term to the scalar potential. In this paper we discuss parametric scale separation that requires an unbound parameter. However, the above statement also applies without such a parameter to any vacuum with numerically large volume and weak coupling. We also notice that the parametric scaling of the volume and 10d string coupling is independent of the compactification dimension. The automatic scale separation for Ricci-flat spaces with harmonic fluxes in a parametric limit was previously argued for in footnote 38 of \cite{Junghans:2020acz}.

For $\langle V\rangle<0$, corresponding to AdS compactifications where supersymmetry implies a dual CFT theory, several works have argued that scale separation is unachievable in certain settings \cite{Gautason:2015tig, Cribiori:2023ihv, Cribiori:2024jwq} or even generically \cite{Lust:2019zwm, Collins:2022nux}. If O-planes are included in explicit AdS/CFT constructions they are usually overpowered by fluxes or branes. Thus, our equation \eqref{eq:scaleseparation} is inapplicable. Moreover, in all examples of the AdS/CFT correspondence that we are aware of, there is a leading curvature contribution $V_R \sim \langle V\rangle$, which as shown in section \ref{sec:leadingterm}, precludes scale separation.\footnote{This holds generically but could potentially be avoided if the internal space has parametric anisotropies.} The argument is essentially that on dimensional grounds $1/L_{\langle V\rangle}^2 \sim V_R \sim R \sim m_{KK}^2$, where $R$ is the internal Ricci scalar. Thus, we highlight, as also shown in \cite{Gautason:2015tig} that scale separation is not achievable with a leading curvature term and even in its absence scale separation would only be required in the presence of a leading O-plane contribution that usually does not arise in the AdS/CFT context.
That curvature obstructs scale separation for M-theory or type II compactifications to four dimensions was previously pointed out in \cite{Gautason:2015tig}, see Appendix \ref{app:Oplenscaleseparation} for a generalization of their argument to arbitrary dimensions.

Our work is inspired by the DGKT construction \cite{DeWolfe:2005uu} and its numerous generalizations (see, e.g., \cite{Camara:2005dc, Ihl:2006pp, Marchesano:2019hfb, Carrasco:2023hta, Tringas:2023vzn, Andriot:2023fss}). The DGKT construction involves massive type IIA flux compactifications with a leading O$6$-plane contribution to the scalar potential. Solutions exist at weak coupling and large volume if the unbounded $F_4$ flux is taken large. As implied by equation \eqref{eq:scaleseparation}, these solutions exhibit automatic parametric scale separation. Various aspects have been discussed, including its uplift to ten dimensions \cite{Acharya:2006ne}, its localization and the presence of Romans mass \cite{Saracco:2012wc, DeLuca:2021mcj}, and concerns regarding the uplift \cite{McOrist:2012yc}. Recently, in \cite{Cribiori:2021djm,VanHemelryck:2024bas} scale-separated solutions in massless type IIA were studied, which allow an M-theory uplift, but have more intricate internal spaces. On the other hand, classical 4D solutions from type IIB with scale separation are not well known. However, attempts have been made in \cite{Caviezel:2009tu, Petrini:2013ika} and further commented on in~\cite{Cribiori:2021djm}.

Scale separation has been intensively explored in dimensions other than four in recent years. Scale-separated 3D solutions have been constructed in type II supergravity with orientifold planes, using internal G2 structure spaces that are Ricci flat or that include metric fluxes. In massive type IIA, several examples with flat G2 structure spaces and O6-planes have been developed \cite{Farakos:2020phe, VanHemelryck:2022ynr, Farakos:2023nms, Farakos:2023wps, Farakos:2025bwf}, exhibiting the same characteristics as DGKT, when the 7d internal space is taken to be isotropic. Interestingly, despite differences in the dimensions of the 3D constructions and DGKT, both exhibit similar behavior when looking at the weak coupling, large volume, and scale separated limit. This suggests that the common fluxes and the leading-order O6-planes play a crucial role in obtaining this limit, as we demonstrate in our paper.
In type IIB, recent constructions with O5-planes have yielded (non-)supersymmetric 3D AdS solutions on G2 spaces with metric fluxes \cite{Arboleya:2024vnp,VanHemelryck:2025qok}, while earlier attempts with calibrated G2 spaces~\cite{Emelin:2021gzx} faced challenges with scale separation due to the quantization of metric fluxes.
Recently, scale-separated AdS$_3 \times S^3$ solutions from 6-dimensional gauged supergravity were constructed without the need for O-planes, see \cite{Proust:2025vmv}. 
Attempts to construct 2D solutions with scale separation from type IIA were explored in \cite{Lust:2020npd}, and recent progress has been made in \cite{Cribiori:2024jwq}.
Effective theories with more than four dimensions and scale separation are not known.
Based on conjectural arguments, as discussed in \cite{Cribiori:2023ihv}, the presence of stable and scale-separated 5D solutions appears to present challenges.
See also attempts to realize scale separation in more holographically motivated constructions in 6D \cite{Apruzzi:2021nle} and 7D \cite{Apruzzi:2019ecr}, which differ from DGKT-like setups.

Here, we systematically investigate whether further scale-separated solutions can exist in type IIA or type IIB across various dimensions, considering different O$p$-planes, $H_3$ flux, and $F_q$ fluxes. Our analysis is based on generic scaling arguments without the calculation of explicit prefactors. We focus on the overall volume and dilaton moduli, leaving detailed examples that include all light fields for future work.

Importantly, equation \eqref{eq:scaleseparation} applies not only to AdS solutions but also to vacua with other values of the scalar potential. For Minkowski vacua ($\langle V \rangle=0$), scale separation is trivial, but moduli stabilization is conjectured to be very difficult \cite{Gautason:2018gln, Andriot:2022yyj}. However, in non-geometric settings there has been lots of recent progress in finding Minkowski vacua with few or no massless fields \cite{Becker:2022hse, Baykara:2023plc, Becker:2023rqi, Becker:2024ijy, Rajaguru:2024emw, Becker:2024ayh}. For $\langle V \rangle>0$ our results hold as well. There are several constructions of dS extrema that use only fluxes, curvature and O-planes, for which the orientifold planes contribute at leading order \cite{Haque:2008jz, Flauger:2008ad, Caviezel:2008tf, Danielsson:2009ff, Caviezel:2009tu, Wrase:2010ew, Danielsson:2010bc, Danielsson:2011au, Shiu:2011zt, VanRiet:2011yc, Danielsson:2012by, Danielsson:2012et, Junghans:2016uvg, Andriot:2016xvq, Junghans:2016abx, Roupec:2018mbn, Kallosh:2018nrk, Andriot:2018ept, Cordova:2018dbb, Cribiori:2019clo, Cordova:2019cvf, Andriot:2019wrs, Andriot:2020wpp, Andriot:2020vlg, Farakos:2020idt, Bena:2020qpa, Andriot:2021rdy, Horer:2024hgy, Chen:2025rkb}. Yet these constructions have not been realized in a trustworthy regime or with scale separation, consistent with equation \eqref{eq:scaleseparation}. In contrast, dS constructions such as KKLT \cite{Kachru:2003aw} or the LVS scenario \cite{Balasubramanian:2005zx} lack a leading orientifold contribution because the orientifold's effect is canceled by $G_3$-flux as in GKP \cite{Giddings:2001yu}. Similarly, the highly scale-separated AdS vacua in \cite{Demirtas:2021nlu, Demirtas:2021ote} fall into the same category as the KKLT and LVS scenarios. Thus, for these later cases our result does not apply.

The outline of this paper is as follows. In section \ref{sec:typeIIcompactifications}, we present our setup, which includes the $D$-dimensional effective theories arising from type II compactifications and the equations of motion relevant for our purposes. We introduce the KK mass approximation that will be used throughout the paper, and present the scaling analysis method, which allows us to examine the parametric behavior of solutions.
In section~\ref{sec:leadingterm}, we examine different terms in the potential that could be leading, such as curvature, fluxes, and local sources, and compare each of them to the KK mass to study potential constraints and universal features regarding scale separation.
In section~\ref{sec:unboundfluxes}, we list possible ways to keep a flux unbounded that could be engineered to examine the parametric behavior of the solutions. Then, applying the scaling analysis again, we present arguments regarding parametric classical solutions, which are related to the dimensionality of the leading O-plane, tadpole cancellation, and the presence of certain fluxes. We discuss the conflict between parametric control and flux quantization, particularly in the context of fluxes that are associated with potential dS solutions.
In section~\ref{sec:classificationscalesep}, we examine the parametric behavior and stability of universal AdS solutions arising from compactification and classify them according to the dimensionality of the O-planes. Lastly, we summarize our findings and discuss future directions in section \ref{sec:Conclusion}.

\section{Type II supergravity compactifications}\label{sec:typeIIcompactifications}
In this section we review the type II supergravity compactifications that we will study in this paper. We discuss compactifications down to a $D$-dimensional maximally symmetric spacetime, keeping track of the dilaton and the overall volume modulus. We include fluxes and localized sources and spell out their contributions to the $D$-dimensional scalar potential. Then we discuss how terms in the scalar potential scale with the dilaton, the overall volume as well as the fluxes and other quantum numbers. Lastly, we study constraints on these scalings that arise from imposing the equations of motion.

\subsection{\texorpdfstring{\boldmath{$D$}-dimensional effective theories}{}}\label{ssec:EFTs}

We construct $D$-dimensional effective theories from type II supergravities with fluxes and smeared sources. The action and equations of motion are reviewed in Appendix~\ref{app:typeII}. In order to obtain a $D$-dimensional effective theory, we start from the 10d type II action in string frame and perform a compactification using the following metric ansatz  
\begin{align}\label{metric1}  
    \text{d}s^2_{10} = \tau^{-2} \,\text{d}s^2_D + \rho\, \text{d}\tilde{s}^2_{10-D}\,,  
\end{align}  
where $\rho$ is the string-frame modulus describing isotropic deformations of the internal space's volume. The $D$-dimensional dilaton is defined as 
\begin{align}\label{Ddilaton}  
    \tau^{D-2} = e^{-2\phi} \rho^{\frac{10-D}{2}} \,, 
    \quad\quad\text{for}\quad\quad
    2< D< 10 \,,
\end{align}  
such that, after this Weyl rescaling of the external spacetime metric, we obtain a $D$-dimensional effective theory in the $D$-dimensional Einstein frame
\begin{equation}\label{eq:SD}
    S_D=M_{Pl}^{D-2}\int d^Dx\sqrt{-g_D}\left(R_D-(D-2)\tau^{-2}(\partial_{\mu}\tau)^2-\frac{10-D}{4}\rho^{-2}(\partial_{\mu}\rho)^2-V\right) \,.
\end{equation}
The $D$-dimensional Planck mass is given by
\begin{equation}\label{eq:MpD}
    M_{Pl}^{D-2}=\frac{1}{2\kappa_{10}^2}\int d^{10-D}y\sqrt{\tilde{g}} = \frac{l_s^{10-D}}{2\kappa_{10}^2} \,,
\end{equation}
where the tilde over the internal space metric $\tilde{g}_{mn}$ in equations \eqref{metric1} and \eqref{eq:MpD} signifies that we are working with a unit-volume internal space in string units. We are setting $M_{Pl}=1$ throughout this paper.

In the presence of internal curvature, $H_3$ flux, RR fluxes and sources the $D$-dimensional scalar potential has the following form\footnote{One can also allow for KK-monopoles and the corresponding OKK-planes. However, as for example discussed in \cite{Junghans:2018gdb, Banlaki:2018ayh} these have the same scaling with $\rho$ and $\tau$ as $V_R$ and thus can be though of as being absorbed into the curvature term $V_R$.}
\begin{equation}\label{Dpotential}
    V
    =
    V_{R}+V_{H}+\sum_{q}V_{F_q}+\sum_{p}V_{Dp/Op}+V_{NS5/ONS5} \,,
\end{equation}
where $q$ and $p$ are the rank of the RR fluxes and sources, respectively. In the democratic formulation, which is used here to write down the Bianchi and flux equations of motion in type II supergravity, we have 
\begin{equation}\label{pq}
\begin{array}{lll}
    \text{IIA: } & \quad q=0,2,4,6,8,10, &\quad  p=0,2,4,6,8,\\
    \text{IIB: } & \quad q=1,3,5,7,9, &\quad  p=1,3,5,7,9.
\end{array}
\end{equation}
Next, we spell out the Bianchi identities and equations of motion for the gauge fields
\begin{align}
    dF_{q}&=H_3\wedge F_{q-2} + \mu_{8-q} J_{q+1} \label{Bianchi1}\,,\\
     d(\star F_{q})&=-H_3\wedge \star F_{q+2} + (-1)^{\frac{q(q-1)}{2}} \mu_{q-2} J_{11-q} \label{flux1}\,,\\
    dH_3&= \mu_{NS5} J^{NS}_{4} \label{Bianchi2} \,,\\
    d(e^{-2\phi}\star H_3)&=-\frac{1}{2}\sum_q\star F_q\wedge F_{q-2} \,.\label{flux2}
\end{align}
Above we have excluded F1-strings as potential sources since we are only interested in maximally symmetric spacetimes with $D>2$. Thus, we consider only sources that extend along all external dimensions. Likewise, all fluxes either need to extend along all external dimensions or none. Given that these two types of fluxes are related to each other by Hodge duality, we will only keep track of fluxes that are entirely extending along the internal space. For the RR fluxes in the democratic formalism we have to impose the duality condition $F_q=(-1)^{\frac{(q-1)(q-2)}{2}} \star F_{10-q}$ to remove redundant fields. This condition means that Bianchi identities in equation \eqref{Bianchi1} and the equations of motion in equation \eqref{flux1} are identical. However, we still spelled them both out because, as just mentioned, we restrict from now on to the fluxes that are purely along internal directions and these have to then satisfy both equations \eqref{Bianchi1} and \eqref{flux1}.

We are interested in static solution for which the lower dimensional equations of motion reduce to the critical point equations $\partial_\tau V=\partial_\rho V=0$.
For later convenience, and only for the purpose of computing the masses, we move to the canonical frame with coordinates $(\tilde{\rho},\tilde{\tau})$, instead of working in the $(\rho,\tau)$ coordinates, by performing the following redefinitions
\begin{align}
\tau\rightarrow e^{\sqrt{\frac{1}{2(D-2)}}\,\tilde{\tau}}\,,
\quad
\quad
\rho\rightarrow e^{\sqrt{\frac{2}{10-D}}\,\tilde{\rho}}\,.
\end{align}
Then, we can compute the mass matrix via $M^2 = \partial_i \partial_j V$, where $i,j = \tilde{\rho}, \tilde{\tau}$. Using the chain rule and re-express the derivatives in terms of the original $(\rho,\tau)$ coordinates we get
\begin{equation}
M^2=
\begin{pmatrix}\label{Hessian}
\partial_{\tilde{\rho}}^2V & \partial_{\tilde{\rho}}\partial_{\tilde{\tau}}V \\
\partial_{\tilde{\rho}}\partial_{\tilde{\tau}}V & \partial_{\tilde{\tau}}^2V
\end{pmatrix}
=
\begin{pmatrix}
\frac{2}{10-D} \rho^2 \partial_{\rho}^2V & \frac{1}{\sqrt{(D-2)(10-D)}}\rho \tau \partial_{\rho}\partial_{\tau}V \\
\frac{1}{\sqrt{(D-2)(10-D)}}\rho \tau \partial_{\rho}\partial_{\tau}V & \frac{1}{2(D-2)}\tau^2\partial_{\tau}^2V
\end{pmatrix} \,.
\end{equation}
The eigenvalues of the above matrix are the two masses squared and they are given by
\be\label{eq:massessquared}
\lambda_\pm  = \frac12 \lp \partial_{\tilde{\rho}}^2V + \partial_{\tilde{\tau}}^2V \pm \sqrt{(\partial_{\tilde{\rho}}^2V - \partial_{\tilde{\tau}}^2V)^2 +4 (\partial_{\tilde{\rho}}\partial_{\tilde{\tau}}V)^2}\rp \,.
\ee

Now we write down the general flux potential for the RR-fields, the NSNS field and the space-filling sources for our metric in equation \eqref{metric1}, see \cite{VanRiet:2011yc},
\begin{align}
    V_{R}&= - \tilde{R}_{10-D}\,\rho^{-1}\tau^{-2} \,,\label{potR} \\
    V_{H}&=\vert \tilde{H}_3\vert^2\rho^{-3}\tau^{-2} \,,\label{potNS} \\
    V_{RR}^q&= \vert \tilde{F}_q\vert^2 \rho^{\frac{10-D-2q}{2}}\tau^{-D} \,,\label{potRR} \\
    V_{Dp/Op}&= T_p\,\rho^{\frac{2p-D-8}{4}}\tau^{-\frac{D+2}{2}} \label{potDBI}
    \,,\\
    V_{NS5/ONS5}&= T_{NS5}\,\rho^{-2}\tau^{-2} \label{potNS5}\,.
\end{align}
Here, we have factored out the modulus dependence from the internal space metric, $g_{mn} = \rho\ \tilde{g}_{mn}$, so that the fluxes and the Ricci scalar are contracted with the tilde metric. The RR and NSNS fluxes should be multiplied by a factor of 1/2 but we assume this is absorbed into the flux squared part of the potential contribution. Generically, there are other moduli in addition to $\rho$ and $\tau$. These turn all of the above prefactors in equations \eqref{potR}-\eqref{potNS5} into functions.

\subsection{General scaling analysis with volume and dilaton}\label{scalinganalysis}

In this work, we do not construct explicit compactifications that keep track of all light fields, but instead we perform a scaling analysis to derive general behaviors based on the overall volume modulus and the dilaton. This captures the universal properties of explicit constructions, such as those in \cite{DeWolfe:2005uu} and \cite{Farakos:2020phe}, but is applicable much more widely to all geometric compactifications.

Examining the Einstein equations \eqref{EinsteinD} and \eqref{EinsteinT}, we see that the dimension of the external space $D$ appears only in the coefficients multiplying the terms in the equations. These coefficients do not enter the scaling analysis, showing that in a $D$-dimensional effective theory, the scaling of fluxes, metrics, and the dilaton is independent of the dimension. For example, in massive type IIA constructions with O6-planes, compactified on 6d \cite{Junghans:2020acz} and 7d \cite{Emelin:2022cac} internal spaces, the scaling analysis of the equations of motion yields the same scalings for all quantities.

On the other hand, in the previous subsection when we were reviewing the dimensional reduction of the scalar potential we found that the dimension $D$ explicitly appears in the scaling analysis in some of the equations \eqref{potR}-\eqref{potNS5}. However, the 10d equations of motion and the equations of motions of the lower dimensional effective theory are equivalent. Thus, the presence of the dimension $D$ in exponents in the lower dimensional equations of motion can be removed by expressing everything in terms of $e^\phi$ and $\rho$ 
\begin{equation}
\begin{array}{rll}\label{eq:Vphi}
    V_{R}&= - \tilde{R}_{10-D}\, \rho^{-1}\tau^{-2} &= e^{\frac{4}{D-2}\phi} \rho^{-\frac{8}{D-2}} \lp - \tilde{R}_{10-D} \rp\,, \\
    V_{H}&=\vert \tilde{H}_3\vert^2\rho^{-3}\tau^{-2} &= e^{\frac{4}{D-2}\phi} \rho^{-\frac{8}{D-2}} \lp \vert \tilde{H}_3\vert^2 \rho^{-2} \rp\,, \\
    V_{RR}^q&= \vert \tilde{F}_q\vert^2 \rho^{\frac{10-D-2q}{2}}\tau^{-D} &= e^{\frac{4}{D-2}\phi} \rho^{-\frac{8}{D-2}} \lp \vert \tilde{F}_q\vert^2 e^{2\phi} \rho^{1-q}\rp\,,\\
    V_{Dp/Op}&= T_p\, \rho^{\frac{2p-D-8}{4}}\tau^{-\frac{D+2}{2}} &= e^{\frac{4}{D-2}\phi} \rho^{-\frac{8}{D-2}} \lp T_p\, e^\phi \rho^{\frac{p-7}{2}} \rp
    \,,\\
    V_{NS5/ONS5}&= T_{NS5}\, \rho^{-2}\tau^{-2} &= e^{\frac{4}{D-2}\phi} \rho^{-\frac{8}{D-2}} \lp T_{NS5}\, \rho^{-1} \rp\,.
\end{array}
\end{equation}
We see that $D$ now only appears in an overall factor of the entire scalar potential. The equations of motion in a vacuum can be written as 
\be
e^{-\frac{4}{D-2}\phi} \rho^{\frac{8}{D-2}}\,\partial_{e^\phi} V = e^{-\frac{4}{D-2}\phi} \rho^{\frac{8}{D-2}} \, \partial_\rho V =0\,.
\ee 
This means that $D$ will again only appear in prefactors in the $D$ dimensional equations of motion but not anymore in any exponent that determines the scaling. Thus, we have proven that the scalings of the terms in the equations of motion, when expressed through $e^\phi$ and $\rho$ are completely independent of the spacetime dimension, allowing for a unified study of the scalings across dimension. 

Since we are interested in scale separation we want to compare the inverse length scale corresponding to the above scalar potential terms with the Kaluza-Klein mass squared. Looking at our metric ansatz in equation \eqref{metric1}, we see that the Kaluza-Klein mass squared scales like $\rho^{-1}$. Since we are also performing a Weyl rescaling by $\tau^{-2}$ all masses squared pick up this additional factor leading to 
\be\label{eq:mKK}
m_{\text{KK}}^2 \sim \rho^{-1} \tau^{-2} = e^{\frac{4}{D-2}\phi} \rho^{-\frac{8}{D-2}}\,.
\ee
Note that this prefactor is exactly the same as the $D$-dependent prefactor in the scalar potential terms in equation \eqref{eq:Vphi}. Thus, in the ratio $\langle V\rangle /m_{\text{KK}}^2$ all terms are independent of $D$, allowing again for a unified treatment of different spacetime dimensions, when we express everything in terms of $e^\phi$ and $\rho$.

\subsection{Scalings with a large quantum number}
Having discussed the scalings with the overall volume and dilaton we now include additional scalings for the curvature and fluxes. We analyze the scalings of the different terms in the scalar potential in terms of a quantum number $N$ that can be taken to infinity. We are focusing on the isotropic breathing mode of the internal space, the $D$-dimensional dilaton, the curvature and the fluxes that thread only internal cycles (after acting on them with the Hodge star if necessary). We make the following generic scaling ans\"atze
\begin{equation}\label{eq:Nscalings}
\begin{split}
    g_{mn}&\sim\rho\sim N^r
    \,,\qquad
    \tau\sim N^t
    \,,\qquad
    e^{\phi} \sim N^d
    \,,\qquad
    \tilde{R}_{10-D}\sim N^c
    \,,\cr
    H_3&\sim N^h
    \,,\qquad\quad\,
    F_q\sim N^{f_q}
    \,,\quad\,\,\,
    T_p\sim N^0
    \,,\qquad\,\,\,
    T_{NS5} \sim N^n \,,
\end{split}
\end{equation}
where the dilaton scaling $d$ is fixed due to equation \eqref{Ddilaton} to be $d=\tfrac{10-D}{4}r+\tfrac{2-D}{2}t$.

In equation \eqref{eq:Nscalings} we have assumed that we include O$p$-planes that arise as fixed point loci of a spatial $\mathbb{Z}_2$ involution. Their number cannot scale with $N$, however, there could be in principle large numbers of D$p$-branes that do scale with $N$. This would amount to a trivial extension of our analysis ($T_p \sim N^p$) but not lead to new interesting results. The reason is that in the presence of a leading curvature term with $\tilde{R}_{10-D} \sim \mathcal{O}(1)$, we find below that parametric scale separation cannot be achieved. If the curvature term is absent or subleading, then the only negative terms in the scalar potential would come from an O$p$-plane or an ONS5-plane. The ONS5-plane or NS5-branes induce a charge that needs to be canceled via equation \eqref{Bianchi2}. In the absence of curvature and metric fluxes this is essentially only possible if these sources cancel each other, leading to $V_{NS5/ONS5}=0$. Thus, we need an O$p$-plane to have a negative term in $V$ or AdS vacua with $V<0$ are not possible. Likewise, dS vacua are forbidden without leading orientifold planes, due to the no-go theorem by Maldacena-Nu\~nez \cite{Maldacena:2000mw}.

Considering the above scalings, the terms in the scalar potential in \eqref{potR} - \eqref{potNS5} scale as follows
\begin{align}
V_R\sim & \,
N^{c-r - 2t} \,, \label{Rscaling}\\
V_H\sim & \, N^{2h -3r - 2t} \,, \label{Hscaling}\\
V^q_{RR}\sim & \, N^{2f_q+\left(\frac{10-D-2q}{2}\right)r-Dt} \sim N^{2f_q+2d-q r-2t} \,, \label{Fnlanescaling}\\
V_{Dp/Op}\sim & \, N^{\frac{2p-D-8}{4}r-\frac{D+2}{2}t} \sim N^{d +\frac{p-9}{2}r-2 t} \,, \label{Oplanescaling}\\
V_{NS5/ONS5}\sim & \, N^{n-2r - 2t}  \,,\label{NSlanescaling}
\end{align}
and we remind the reader that $D$ denotes the dimension of the external spacetime and $d$ the scaling of the dilaton with respect to $N$, see equation \eqref{eq:Nscalings}.

The flux equations of motions and Bianchi identities are a little bit more involved, and we discuss them and their scalings in detail in Section \ref{sec:unboundfluxes}. The remaining equations are Einstein equations in equations \eqref{EinsteinD}, \eqref{EinsteinT} as well as the dilaton equation. However, these are equivalent to the $D$-dimensional equations of motion that in a vacuum are simply given by $\partial_\tau V=\partial_\rho V=0$ or alternatively by $\partial_{e^\phi} V=\partial_\rho V=0$. If we look at the derivative with $\rho$, then all the terms in $\rho \partial_\rho V(\tau,\rho)=0$, have exactly the same scalings as in \eqref{Rscaling}-\eqref{NSlanescaling}, as can be seen from the minimization of the potential
\begin{equation}\label{eq:MinVrhogeneral}
    0= -V_{R}
    -3V_H
    +\frac{1}{2}(10-D-2q)\sum_q V_{RR}^q
    +\frac{1}{4}(2p-D-8)V_{Dp/Op} 
    -2V_{NS5/ONS5}\,.
\end{equation}
Similarly, the scalings of the terms in $\tau \partial_\tau V=0$ and $e^\phi \partial_{e^\phi}V=0$ have the same scalings as those in \eqref{Rscaling}–\eqref{NSlanescaling}, as can be seen from the structure of the equations of motion 
\begin{equation}\label{eq:MinVtaugeneral}
    0=2V_R
    +2V_H
    +D\sum_q V_{RR}^q
    +\frac{1}{2}(D+2)V_{Dp/Op}
    +2V_{NS5/ONS5}\,.
\end{equation}
Thus, imposing that certain terms in the equations of motion scale in the same way is identical to imposing that they scale in the same way in the scalar potential itself.

\section{Analysis of the leading scalar potential terms}\label{sec:leadingterm}

In this section, we present some interesting results regarding scale separation when different types of terms are leading in the scalar potential. We recall from above that these results are generically applicable to all geometric compactifications to arbitrary $D>2$. The idea is as follows: The scalar potential $V$ in equation \eqref{Dpotential} is the sum of several terms. However, not all terms are necessarily non-zero or even if they are non-zero they could be subleading compared to others as we will see below. We now discuss in turn different leading terms in the scalar potential and compare them with $m_{\text{KK}}^2 \sim \rho^{-1} \tau^{-2}$ from equation \eqref{eq:mKK}. 

\paragraph{Leading curvature term:} 
\hspace{0.5cm}\\
In the AdS/CFT context all examples we are familiar with, have an internal space that has a non-zero curvature and the term $V_R$ is a leading term in the scalar potential. This means that the inverse length scale associated to our scalar potential vev $\langle V\rangle$ satisfies $1/L_{\langle V\rangle}^2\sim V_R =- \tilde{R}_{10-D}\,\rho^{-1}\tau^{-2}$, where we used equation \eqref{potR}. Thus, one generically has
\begin{align}\label{eq:VrovermKK}
    \frac{V_{R}}{m^2_{\text{KK}}} \sim \frac{-\tilde{R}_{10-D} \rho^{-1} \tau^{-2}}{\rho^{-1} \tau^{-2}}  
    \sim -\tilde{R}_{10-D} \,.
\end{align}
Note that $\tilde{R}_{10-D}$ is \emph{not} the curvature of the internal space but the curvature of the unit-volume metric $\tilde{g}_{mn}$, which would be a simple constant for example for $S^5$ or $\mathbb{RP}^5$. More generically, there are spaces where the curvature term scales $\tilde{R}_{10-D}\sim N^c$. This arises, for example, if we T-dualize $H_3$ flux into metric flux. However, in that case and all others that we are aware of, the exponent $c>0$, which means that $\tilde{R}_{10-D}$ can never be small. Whenever this is the case this means that no matter what the volume of the internal space or the value of the dilaton, we can never get scale separation! It would be interesting to pursue this further. On the one hand, one could try to prove that curvature is always required in the AdS/CFT context and therefore scale separation is always impossible, as was conjectured in \cite{Lust:2019zwm, Collins:2022nux}. Alternatively, one can also try to find the CFT dual theory for the DGKT construction \cite{DeWolfe:2005uu}\footnote{It was recently argued in \cite{Montero:2024qtz} that the DGKT AdS vacua might be all non-supersymmetric so that no dual CFT theory would exist because non-supersymmetric AdS vacua are expected to decay \cite{Ooguri:2016pdq}.} or the many generalizations thereof that also give rise to AdS$_3$ vacua \cite{Farakos:2020phe,Farakos:2023nms,Farakos:2023wps,Farakos:2025bwf} from type IIA, which should have CFT$_2$ duals.

Note that some versions of DGKT involve formal T-dualities of the internal space that turn the $H_3$ flux into curvature \cite{Cribiori:2021djm}.
The corresponding spaces are not necessarily Ricci-flat, but the Ricci scalar can still vanish, or anisotropies can lead to internal spaces with cycles that scale parametrically differently from one another. This modifies the above analysis, as seen for example in AdS$_3$ constructions from type IIB with anisotropic internal space and non-vanishing curvature \cite{Arboleya:2024vnp, VanHemelryck:2025qok}, which can lead to scale separation.
Similar AdS$_3$ constructions, but with isotropic internal space and non-vanishing curvature, do not achieve scale separation, see \cite{Emelin:2021gzx}. Likewise, isotropic internal spaces with curvature prevent scale separation in AdS$_4$ constructions \cite{Tsimpis:2012tu, Font:2019uva}.

\paragraph{Leading \boldmath{$H_3$} flux term:} 
\hspace{0.5cm}\\
Next, we consider a leading term $V_H$ from equation \eqref{potNS} and compare it to the KK-mass scale to find
\begin{align}
    \frac{V_{H}}{m^2_{\text{KK}}}\sim \frac{\vert \tilde{H}_3\vert^2\rho^{-3}\tau^{-2}}{\rho^{-1}\tau^{-2}}
    \sim
    \vert \tilde{H}_3\vert^2\rho^{-2} \,.
\end{align}
This means that scale separation is automatic if we are at large volume, $\rho\gg1$, but it is independent of the value of $\tau$. There is one caveat because in principle $H_3\sim N^h$ can scale and become large in a given solution as indicated in equation \eqref{eq:Nscalings}. Taking the $\rho$ scaling $\rho \sim N^r$ into account we find scale separation for large $N$ solutions that satisfy $h<r$.

\paragraph{Leading \boldmath{$F_q$} flux term:} 
\hspace{0.5cm}\\
For a leading term that arises from an RR flux, we now use equation \eqref{eq:Vphi} to get an answer that is independent of the spacetime dimension $D$
\begin{align}\label{eq:Vrrovermkk}
    \frac{V^q_{RR}}{m^2_{\text{KK}}}\sim \frac{\rho^{-1}\tau^{-2} \lp \vert \tilde{F}_q\vert^2 e^{2\phi} \rho^{1-q}\rp}{\rho^{-1}\tau^{-2}} \sim \vert \tilde{F}_q\vert^2 e^{2\phi} \rho^{1-q} \,.
\end{align}
For the special case of massive type IIA with $q=0$, a small volume and weak coupling seems desirable. Otherwise, for $q\geq1$, we see that again a weak coupling and/or large volume limit seems to help with scale separation. The flux quanta might also scale with $N$ as described in equation \eqref{eq:Nscalings}. In that case we have scale separation if $2f_q+2d-(q-1) r<0$. Note, that in any large $N$ limit we have $f_q>0$ due to flux quantization. $d$ is the scaling of the dilaton which can have either sign and is related to the scaling of $\tau$ and $\rho$ via equation~\eqref{Ddilaton}.

The above relation in equation \eqref{eq:Vrrovermkk} is also valid off-shell and was used recently in~\cite{Andriot:2025cyi} to show that scale separation is achieved for rolling solutions without the presence of orientifolds.

\paragraph{Leading O\boldmath{$p$}-plane term:} 
\hspace{0.5cm}\\
So far, all known constructions with parametric scale separation involve O-planes, suggesting that their presence is an essential ingredient for achieving scale separation.
As we have explained above, it is reasonable for the O-plane contribution $V_{Op}$ to be leading since in the absence of curvature this would enable the existence of AdS vacua and for dS vacua O-planes provide a way to circumvent the Maldacena-Nu\~nez no-go theorem~\cite{Maldacena:2000mw}. Comparing a leading term $V_{Dp/Op}$ with the KK-mass we find from equation~\eqref{eq:Vphi}
\begin{align}\label{eq:VOpovermKK}
    \frac{V_{Dp/Op}}{m^2_{\text{KK}}}
    \sim
    \frac{\rho^{-1}\tau^{-2}\lp T_p\, e^\phi \rho^{\frac{p-7}{2}}\rp}{\rho^{-1}\tau^{-2}}
    =
   T_p\, e^\phi \rho^{\frac{p-7}{2}} \,.
\end{align}
So, we see that if D-branes or O-planes are a leading contribution to the scalar potential then scale separation is automatically achieved for weak coupling and/or large volume if $p<7$. This means scale-separation comes for free in the supergravity regime! In the case of D-branes we are however having a positive contribution to the scalar potential. Looking at equations \eqref{potR}-\eqref{potNS5}, we see that all terms in the scalar potential scale with a negative power of $\tau$ and thus no extrema exist unless we have at least one negative term in the scalar potential. Thus, the prospect of a leading order negative term from an O-plane seems much more relevant than the D-brane case. Also, for D-branes the prefactor $T_p$ is proportional to the number of branes and if this number is large then this could prevent scale separation even at large volume and weak coupling.

\paragraph{Leading NS5 or ONS5 source term:} 
\hspace{0.5cm}\\
Lastly, we look at the case where the term $V_{NS5/ONS5}$ in equation \eqref{potNS5} provides a leading order contribution to the scalar potential
\begin{align}
    \frac{V_{NS5/ONS5}}{m^2_{\text{KK}}}
    \sim
    \frac{T_{NS5}\,\rho^{-2}\tau^{-2}}{\rho^{-1}\tau^{-2}}
    =
   T_{NS5}\,\rho^{-1}\,.
\end{align}
We see that also in this case a large volume would lead to scale separation. For the ONS5 case we would also automatically have the needed negative term in the scalar potential. It would be interesting to explore this further but we refrain from doing so in this paper and restrict below to studying examples without NS sources.\\

In general, in order to have an actual minimum of the scalar potential requires that there are several leading order terms in the scalar potential. Specifically, looking at equations \eqref{potR}-\eqref{potNS5} we see that the negative powers of $\tau$ means that for AdS vacua we need at least two leading terms while for dS vacua we need at least three leading terms. In the case that multiple of the terms discussed above are leading then multiple of the conclusions we reached above are simultaneously applicable to such vacua. Most notable to us are the universal insights that a leading curvature term prevents scale separation while a leading O-plane contribution implies scale separation for either weak coupling and/or large volume solutions.

If we have a parameter $N$, like for example an unbound flux that we can make large, then we can study the scaling with this parameter and we can expand the scalar potential to get the leading order ($LO$) and next to leading orders ($NLO$) 
\begin{equation}
V = V_{LO} N^a + V_{NLO} N^b + \dots \,, \quad \text{for}\quad a > b \,.
\end{equation}
While $V_{LO}\neq 0$ sets the inverse length scale and is what enters into our discussion above, the subleading terms can also play an important role. They could stabilize further moduli as in DGKT \cite{DeWolfe:2005uu}, where they also encode corrections to the smeared solution that arise when we localize the O6-planes \cite{Junghans:2020acz, Marchesano:2020qvg, Emelin:2022cac,Andriot:2023fss,Emelin:2024vug, VanHemelryck:2024bas}.

\section{\texorpdfstring{Unbounded fluxes and some interesting scaling results}{Unbounded fluxes and some interesting scaling results}}\label{sec:unboundfluxes}

To discuss the parametric behavior in our constructions, we need an unbounded parameter, typically a flux. When all quantities are expressed in terms of this parameter and it is taken to be large, we obtained controlled parametric results. For the flux to be unbounded, the equations of motion must permit this and must not impose constraints on its magnitude. The Einstein equations and the dilaton equation contain the volume and string coupling that can scale as well and compensate to some extent for different scalings of the fluxes. So, in the next two subsections we first focus on restrictions arising from the Bianchi identities and flux equations of motion. Then we will use the remaining equations of motion to derive interesting scaling relationships in the last three subsections.

\subsection{Trivially unbounded fluxes}
So far, essentially two methods have been developed in the literature that leave a flux unconstrained. In this subsection we discuss what we call a trivially unbounded flux, which describes a flux, either $H_3$ or $F_q$, that does not appear in any of the Bianchi identities or the flux equations of motion. 

First, for a flux quanta to exist the internal compactification space needs to have a corresponding cycle that is threaded by the flux. We expand the flux in terms of harmonic forms. So, let $A_q$ be a harmonic $q$-form that could either correspond to an RR-flux $F_q$ or to $H_3$. Since $A_q$ is harmonic, we have $dA_q = d(\star A_q)=0$. In the cases we are interested in below we will have smeared sources and a constant dilaton so that $d(e^{-2\phi} \star A_q)=0$ as well. This means in the relevant equations \eqref{Bianchi1}-\eqref{flux2} the left-hand-sides are all equal to zero for either $A_q= F_q$ or $A_q= H_3$. 

Next we look at the right-hand-side of the equations, which involve wedge products of the fluxes and a potential action with the Hodge star. If all of these wedge products involving the particular flux $A_q$ do vanish, then it is trivially unbounded. There can be different reasons for this to happen. For example, it is possible that wedge products of $A_q$ with other fluxes give a $q+n$ form and $q+n>10-D$. This means the resulting form degree exceeds the dimension of the internal space and the wedge product must vanish identically. Even if the resulting form degree is allowed ($n + p \le 10-D$), it vanishes if it corresponds to a trivial cohomology class. For example, on a Calabi--Yau threefold, the wedge product of a harmonic $3$-form and a harmonic $2$-form produces a harmonic $5$-form. However, Calabi--Yau threefolds do not admit nontrivial $5$-cycles or corresponding harmonic $5$-forms. Consequently, equations involving wedge products of fluxes such as
\begin{equation}
H_3 \wedge \star_6 F_4 = 0\,, \quad\quad H_3 \wedge F_2 = 0 \,,
\end{equation}
vanish trivially, imposing no additional constraints on the fluxes. 

Lastly, we can also choose other fluxes so that one particular flux $A_q$ does not appear in the Bianchi identities or equations of motion. For example, on the right-hand-side of equation \eqref{flux2} we have
\be
\sum_s \star F_s \w F_{s-2} = \ldots + \star F_{q+2} \w A_q + \star A_q \w F_{q-2} + \ldots\,.
\ee
We can remove the $A_q=F_q$ dependence completely by choosing appropriate $F_{q+2}$ and $F_{q-2}$, with $F_{q+2} = F_{q-2}=0$ being an obvious example.

Thus, we have seen that it is possible to have setups where a certain RR or NSNS flux does not appear at all in the Bianchi identities and equations of motion for the fluxes. The prime example where this happens is the DGKT setup with an unbounded $F_4$ flux \cite{DeWolfe:2005uu}.

\subsection{Non-trivially unbounded fluxes}
There are other cases in which we can have an unbound flux quantum number. They come in many different disguises, but in the end they boil down to a particular flux quantum number not appearing at all in the equations of motion for the fluxes and the Bianchi identities. The distinguishing feature when compared to the previously discussed case is that the corresponding flux does appear in the flux equations of motion and/or Bianchi identities but the particular unbound flux quantum number does not. 

The most generic unbound flux situation that one could encounter is one in which a particular flux quantum number does not appear in the Bianchi identities in equations \eqref{Bianchi1} and \eqref{Bianchi2}. The reason is that in those equations we have only quantized fluxes and quantized source charges. Since none of these can become parametrically small, no flux quantum number appearing explicitly in these equations can become parametrically large either. However, a parametrically large flux quanta could in principle appear in the flux equations of motion that involve the dilaton and metric via the Hodge star, see equations \eqref{flux1} and \eqref{flux2}. The scalings of the dilaton and metric components could then in principle allow for a flux quantum number going to infinity but we are not aware of such a solution in the existing literature. So, the cases we discuss below have the unbound flux quantum number neither appearing in the Bianchi identities nor in the flux equations of motion.

\paragraph{Vanishing wedge product:}

It is instructive to discuss a variety of examples that have appeared in the literature. Let us first start with an example from \cite{Farakos:2025bwf}. There the authors had non-vanishing $H_3\neq0$ and $F_4\neq0$ fluxes. They studied a compactification on a seven dimensional space so that $H_3 \w F_4$ is proportional to the volume form of the compactification space. However, they chose the fluxes so that $H_3 \w F_4=0$. This example trivially generalizes to any other $F_q$ flux so we generalize here; however, if $q+3$ is not the dimension of the compactification space, then a similar analysis would apply to each independent $q+3$ harmonic form. 

Expanding the fluxes in a basis of harmonic $n$-forms $\omega_i^{(n)}$ as $H_3 = \sum_i h_i \omega_i^{(3)}$ and $F_q = f_i \omega_i^{(q)}$, one has to ensure that
\begin{align}\label{fluxcancel}
    \sum_i h_i f_i=0\,.
\end{align}
The authors of \cite{Farakos:2025bwf} chose all $h_i$ and $f_i$ to be non-zero while satisfying the above equation. By a change of basis this is equivalent to what was used in \cite{Farakos:2023nms}, namely a choice where for any given $i$ either we have $h_i=0$ or $f_i=0$. 

In the presence of O$(6-q)$ planes, the above choices require the presence of D$(6-q)$-branes to cancel the charge
\begin{equation}
0 = H_3 \wedge F_{q} + \mu_{D(6-q)} J_{q+3}+ \mu_{O(6-q)} J_{q+3}= \mu_{D(6-q)} J_{q+3}+ \mu_{O(6-q)} J_{q+3} \,.
\end{equation}
However, it is in principle also possible to choose the flux quantum numbers in equation \eqref{fluxcancel} so that they do not give zero but a finite number that cancels the O-planes directly, which leads us to our next case study.
    
\paragraph{Decomposing a flux:}

We can extend the above analysis by decomposing the RR flux into two components,
\begin{equation}\label{crankup}
    H_3\wedge F_q=H_3\wedge F_{q,A}+H_3\wedge F_{q,B}\,,
\end{equation}
such that one part cancels against the source charge, resulting in $F_{q,B} \sim N^0$, while the other component, $F_{q,A} \sim N^{f_A}$, cancels trivially via methods discussed above
\begin{align}
    0=&H_3\wedge F_{q,B}+\mu_{6-q}J_{q+1} \,,\label{cc}\\
    0=&H_3\wedge F_{q,A} \,.
\end{align}
This is a simple extension of the case above that does not require D-branes.

As expected, the decomposition of the flux breaks the relevant RR potential term into three terms, $V_{RR}^q\equiv V_{RR,AA}^q+V_{RR,AB}^q+V_{RR,BB}^q$, with different scaling behaviors 
\begin{equation}\label{crankpot}
   V_{RR,AA} \sim N^{f_{q,A}}V_{RR,AB}\sim N^{2f_{q,A}}V_{RR,BB} \,.
\end{equation}
This will create three different scales in the effective theory since
\begin{equation}
    V_{RR,AA}>V_{RR,AB}>V_{RR,BB} \,,\quad\text{for}\quad N\gg 1\,,
\end{equation}
with the terms $V_{RR,AB}$ and $V_{RR,BB}$ being subleading in the scalar potential. This is actually not possible for a homogeneously scaling space as we prove below in subsection~\ref{ssec:weaklargelimit}. There we show that the flux that cancels the leading O-plane, $F_{q,B} \sim N^0$, in this case, gives necessarily rise to a leading term in the scalar potential, $V_{RR,BB}$. Thus this case cannot arise unless the internal space has cycles with parametrically different sizes. Examples with parametrically large volume, weak coupling and parametrically different internal cycles are discussed for example in \cite{Cribiori:2021djm,Farakos:2023nms,Tringas:2023vzn,Farakos:2023wps,Arboleya:2024vnp,VanHemelryck:2025qok,Farakos:2025bwf}.

\paragraph{Non-closed forms:}  

Another seemingly different way to cancel the tadpole of an O-plane while leaving some fluxes unconstrained is by considering the magnetic flux appearing in the Bianchi identities \eqref{Bianchi1} to be partially non-closed $dF_q \neq 0$. Simply canceling the charge by considering the entire $dF_q$ contribution would bound the flux~$F_q$, preventing parametric scale separation, as discussed in \cite{Emelin:2021gzx}. Therefore, part of the flux should contribute to the tadpole cancellation, while other components can be expanded in closed forms that do not appear in the Bianchi identity and remain unconstrained, see for example \cite{VanHemelryck:2025qok}. Generically we have 
\begin{align}\label{eq:BianchinotclosedFq}
    dF_{q}=dF_{q}^{\rm (closed)}+dF_{q}^{\rm (non-closed)} =dF_{q}^{\rm (non-closed)} =\mu_{8-q} J_{q+1}\,.
\end{align}
This means necessarily that we are not dealing anymore with harmonic forms. However, this case is often related to the above by a formal T-duality that turns the $H_3$ flux into so-called metric fluxes. For compactification spaces with a special group structure one can often restrict to forms that are invariant under the group action, which provides a consistent truncation. One can then expand $F_{q}^{\rm (closed)}$ in closed $q$-forms that are invariant under the group action and try to find solutions with the corresponding unconstrained flux quantum numbers.

The Bianchi identity in equation \eqref{eq:BianchinotclosedFq} could additionally involve a term $H_3 \w F_{q-2}$. However, if $dF_{q}^{\rm (non-closed)}$ contributes to a particular $q+1$ form, then this form is necessarily exact. So we would not have a flux contribution from the $H_3 \w F_{q-2}$ term. However, they can contribute to different $q+1$ forms as for example in \cite{Cribiori:2021djm}. There the DGKT setup was formally T-dualized and some non-closed $F_2^{\rm (non-closed)}$ flux was canceling the tadpole, while some closed $F_2^{\rm (closed)}$ flux piece led to large volume and weak coupling.

\subsection{\texorpdfstring{No scaling of the \boldmath{$H_3$} and \boldmath{$F_{6-p}$} flux quanta}{}}\label{ssec:noscaling}
In addition to the Bianchi identity we have to also impose the equations of motion in a vacuum, $\partial_\tau V = \partial_\rho V=0$. We will restrict our discussion to cases without curvature $V_R=0$, no NS sources and an O$p$-plane. In that case we now prove that the $H_3$ and $F_{6-p}$ fluxes cannot scale at all. As we explained above, this does not follow from the Bianchi identity in equation \eqref{Bianchi1}, since the flux piece that we take to be large could simply not appear in the Bianchi identity. However, using the equation of motion $\partial_\tau V=0$, we can exclude this and show that $h=f_{6-p}=0$ in equation \eqref{eq:Nscalings}. 

For $V = V_H + \sum_q V^q_{RR} + V_{Op}$, see equations \eqref{eq:Vphi}, we can solve $\partial_{\tau} V =0$ in terms of $T_p$ to find
\begin{align}\label{eq:Tp}
    T_p=-\frac{4\vert\tilde{H}_3\vert e^{-\phi}\rho^{\frac{3-p}{2}}+2D\sum_q\vert \tilde{F}_q\vert^2e^{\phi}\rho^{\frac{9-p-2q}{2}}}{D+2} \,,
\end{align}
where we used $\tau^{D-2} = e^{-2\phi} \rho^{\frac{10-D}{2}}$ from equation \eqref{Ddilaton}. Since all fluxes thread the internal space, we have $\vert \tilde{H}_3\vert^2, \vert \tilde{F}_q\vert^2>0$. This means that $T_p$ is negative as required for an orientifold plane. In any large $N$ limit where the fluxes as well as $\rho$ and $\tau$ scale with a certain power of $N$ there will be one or more leading terms in equation \eqref{eq:Tp}. Since the leading terms are necessarily all positive definite they cannot cancel each other (but subleading terms could). This means that the leading order term(s) cannot scale at all with $N$ since the left-hand-side is the O$p$-plane tension that cannot scale with $N$. Let us first consider the case where the terms involving $H_3$ and $F_{6-p}$ are both leading terms so that
\begin{equation}\label{eq:HwFnoscale}
\begin{split}
    \vert\tilde{H}_3\vert e^{-\phi}\rho^{\frac{3-p}{2}}&\sim N^{2h-d+\frac{3-p}{2}r}\sim N^0 \,, \\
    \vert \tilde{F}_{6-p}\vert^2e^{\phi}\rho^{\frac{9-p-2(6-p)}{2}}&\sim N^{2f_{6-p}+d-\frac{3-p}{2}r}\sim N^0 \,.
\end{split}
\end{equation}
This implies that $h=-f_{6-p}$, which implies that the leading terms of both $H_3$ and $F_{6-p}$ have to appear in $H_3 \wedge F_{6-p}$ in order to get an $N^0$ scaling that can cancel the O$p$-plane charge in the Bianchi identity \eqref{Bianchi1}. If that were not the case then the O$p$-plane charge could not be canceled in the large $N$ limit since $H_3 \wedge F_{6-p}$ would go to zero. Likewise, if not both $H_3$ and $F_{6-p}$ do appear at leading order $N^0$ in equation \eqref{eq:Tp}, then their contribution $H_3 \wedge F_{6-p}$ in the Bianchi identity would go to zero for large $N$ and they again could not cancel the O$p$-plane charge. Since $H_3 \sim N^h$ and $F_{6-p} \sim N^{f_{6-p}}$ and all fluxes are quantized we noticed above that $h, f_{6-p}\geq 0$. With $h=-f_{6-p}$ this completes the proof that $h=f_{6-p}=0$. 

Note that the above proof assumed a single O$p$-plane. However, it extends to the case of several different O-planes in which case it applies to the O$p$-plane with the largest $p$. This follows from the fact that the scalar potential terms $V_{Op}\sim \rho^{\frac{2p-D-8}{4}} \tau^{-\frac{D+2}{2}}$ are at large volume $\rho\gg1$ dominated by the largest $p$ value. This then sets the overall scale of the scale potential and the other $V_{Op'}$ with $p'<p$ are subdominant.

\subsection{Weak coupling and large volume limit}\label{ssec:weaklargelimit}
Here we discuss one more scaling result that follows from the observation above in the scaling relations \eqref{eq:HwFnoscale}. We can directly read off from $h=f_{6-p}=0$ that 
\be
d=\frac{3-p}{2} r \,.
\ee
In the large $N$ limit, a small ten dimensional dilaton $e^\phi \sim N^d$ requires that $d<0$ and a large internal volume $\rho^{\frac{10-D}{2}} \sim \rho^{\frac{10-D}{2}r}$ requires that $r>0$. We see that this is achievable if and only if $p>3$. So, in our type of setup we cannot get parametrically controlled solutions if we do not include O$p$-planes with $p>3$.

Note that the above $d<0$ and $r>0$ automatically imply that the lower D-dimensional dilaton $\tau$ is small. This follows from its definition $\tau^{D-2} = e^{-2\phi} \rho^{\frac{10-D}{2}}$. We actually see that a large $\tau \sim N^t$ is achieved whenever
\be\label{eq:trrelation}
t =\frac{2(p+2)-D}{2 (D-2)} r>0\,.
\ee
This follows automatically from $r>0$, since we are restricting to $D\geq 3$ with spacetime filling O$p$-planes, i.e. $p+1\geq D$.

In subsection \ref{ssec:noscaling} we showed that it is necessary that $H_3$ and $F_{6-p}$ are leading terms in equation \eqref{eq:Tp}. This actually implies that in the large $N$ limit the three terms, $V_{Op}, V_H$ and $V_{RR}^{6-p}$ in the scalar potential are necessarily leading. Using the relation in equation \eqref{eq:trrelation} we find that the scalar potential and these three leading terms scale as
\be\label{eq:Vlead}
V \sim V_{Op} \sim V_H \sim V_{RR}^{6-p} \sim \rho^{-\frac{2(D+p-1)}{D-2}} \,.
\ee

\subsection{The scaling of other RR fluxes}\label{ssec:NoquantizedRRforsmallq}
Above in subsection \ref{ssec:noscaling} we have shown that with a leading O$p$-plane and a large parameter $N$, the $H_3$ flux and the $F_{6-p}$ flux cannot scale with $N$. This has further implications for other RR-fluxes: Let us calculate the ratio of $V^q_{RR}$ and $V^{6-p}_{RR}$ using equation \eqref{eq:Vphi} and the $N$ scalings in equation \eqref{eq:Nscalings}
\be
\frac{V^q_{RR}}{V^{6-p}_{RR}} \sim N^{2 f_q-2f_{6-p}+(6-p-q)r}\,.
\ee
We saw above that $V_{RR}^{6-p}$ is a leading term in the potential, so either $V^q_{RR}$ is subleading and we can neglect it for most of our arguments below or it is also leading in which case the above ratio has to order one, i.e., scale like $N^0$. Using that $f_{6-p}=0$ the latter implies that
\begin{align}\label{fluxbehavior}
    f_q =\frac{q-(6-p)}{2} r \,,
\end{align}
which consequently means that for $q < 6-p$ and $r > 0$, the relevant flux $F_q$ would become infinitesimally small in the large $N$ limit, $F_{q} \sim N^{f_q} \to 0$. This is inconsistent with flux quantization. Thus, our assumption above that an RR flux $F_q$ is equally important as the $F_{6-p}$ flux is not possible for $q < 6-p$ and all such fluxes are necessarily subdominant in a parametric limit. Rephrasing this, we have shown that the RR flux $F_q$ that appears in the Bianchi identity and cancels the O-plane charge necessarily has the smallest $q$ from all the RR fluxes that appear at leading order in the scalar potential. 

\subsection{No parametrically controlled dS vacua}\label{ssec:nodS}
Interestingly, the combination $q -(6-p)$, which determines the scaling of the extra RR flux in equation \eqref{fluxbehavior} above, also appears as a coefficient in the vacuum expectation value of the scalar potential and dictates whether a term contributes positively or negatively.
Before proceeding, let us focus on the three leading terms from equation~\eqref{eq:Vlead}
\begin{equation}\label{eq:Minkowski}
   V = V_H + V_{Op} + V_{RR}^{6-p} \,,    
\end{equation}
where $V_{RR}^{6-p}$ corresponds to the potential term of the RR flux involved in the tadpole cancellation together with $H_3$. This setup includes the GKP construction \cite{Giddings:2001yu}, its generalization to other dimensions \cite{Blaback:2010sj}, and further compactifications built upon them. 
Minimizing the scalar potential with respect to the dilaton and volume we find
\begin{align}
    V^{6-p}_{RR}=V_H\,,\quad
    V_{Op}=-2V_H\,.
\end{align}
We see that the conditions we get are independent of the dimensions of the theory or the type of O$p$-plane. It is straightforward to verify that, by plugging this into the potential in \eqref{eq:Minkowski}, we always obtain a vanishing cosmological constant $\langle V \rangle = 0$, since the equations of motion require the above three terms to cancel each other exactly.

The exact cancellation among the original terms, which lead to Minkowski solutions, still holds in the presence of additional terms in the scalar potential. However, the Minkowski solutions are modified due to the inclusion of these extra terms.
Let us take the scalar potential to be
\be
V = V_H + V_{Op} + V_{RR}^{6-p} + \sum_{q\neq 6-p} V_{RR}^{q}\,.
\ee
Then we can solve $\partial_\tau V = \partial_\rho V =0$ and use these two equations of motion to fix $V_H$ and $V_{Op}$ in terms of the $V_{RR}^q$, this leads to
\begin{equation}\label{eq:VAdS}
    \langle V\rangle=-\frac{1}{2}\sum_{q\neq 6-p} \frac{(D-2)(q-(6-p))}{D+p-1}\,V_{RR}^q \,.
\end{equation}
We see that $V_{RR}^{6-p}$ does not contribute in equation \eqref{eq:VAdS} since it combines with $V_H$ to cancel the negative $V_{Op}$.
We began constructing vacua by including the essential harmonic fluxes required to cancel the tadpole, which led to Minkowski solutions. After introducing additional fluxes, the resulting expression for the vacuum expectation value allows us to explore the possibility of dS and AdS solutions.

Let us now discuss the possibility of obtaining dS solutions with parametric control in these setups.
The above expression for the vacuum expectation value shows that parametrically controlled dS vacua could, in principle, be obtained if $-(q - (6 - p)) > 0$, which holds whenever $q < 6 - p$.
However, we have just shown in \eqref{fluxbehavior} that this is incompatible with flux quantization in a large $N$ limit. 
More specifically, it can be shown that fluxes contributing positively to the vacuum expectation value appear only in constructions with O2-, O3-, and O4-planes, see Table \ref{tablevacuum}, and the presence of such quantized fluxes does spoil parametric control. This holds for any maximally symmetric solution that includes these fluxes and leading O-planes.

However, once the Bianchi identities are taken into account, further restrictions arise for some setups, and the possibilities of obtaining dS solutions in Table \ref{tablevacuum} become more limited.
For O4-planes, the Romans mass $F_0$ must be set to zero to satisfy the Bianchi identity $dF_0=0$ since it has be odd under the spatial part of the orientifold projection, in agreement with the no-go theorems in \cite{Andriot:2022xjh}. Therefore, the positive contribution from the Romans mass in O4-plane solutions should be set to zero in Table~\ref{tablevacuum}, and only AdS solutions can be obtained, whose parametric limit we study later.

Additionally, making use of the relevant Bianchi identity, $dF_5 = H_3 \wedge F_3 + \mu_3 J_6$, one can also exclude dS solutions for O3-planes in $D = 4$. The reason is that the Bianchi identity provides an extra constraint that implies that $\langle V \rangle \leq 0$. The only positive contribution to the vacuum energy, $V_{RR}^1$, cancels out, while the other terms give only negative contributions. Even if $F_5$ has a non-closed piece, $F_5=F_5^{(closed)}+F_5^{(non-closed)}$, the integral of a potentially positive term $\sim dF_5$ still vanishes.
It has been shown in \cite{Andriot:2022xjh} that in the case where $p = D - 1$, the integral $\int_{10-D}dF_{8-p}=0$, and dS solutions are highly constrained.
However, a $D = 3$ setup with O3-planes does not fall into this case and is not excluded from admitting dS solutions by any no-go theorem so far.
In this case, the integral of the $F_5$ term does not vanish and can contribute positively to the vacuum energy. 
Since this term is involved in the Bianchi identity and the cancellation of the tadpole, it is bounded and should not scale. 
This means that the quantities appearing in $dF_5^{(non-closed)} \sim f_5\, \omega^{(6)}$ either do not scale at all, or their scalings are inversely proportional so that the overall scaling cancels.
The first case breaks the scaling symmetry for isotropic spaces, while the second would be in conflict with flux quantization. 

Lastly, for O2-planes in $D=3$, as one can see in Table \ref{tablevacuum}, there are two positive contributions from $F_0$ and $F_2$, but the latter cancels out after applying the Bianchi identity, $dF_6=H_3\wedge F_4+\mu_{2}J_7$. 
Thus, the Romans mass remains the only positive contribution. To keep the Romans mass non-zero and satisfy the Bianchi identity $dF_2 = H_3\wedge F_0$, either the $H_3$ flux vanishes and $F_6$ is non-closed, or $F_2$ is a non-closed form and satisfied the Bianchi. In any case, non-closed forms must be taken into account.
The presence of non-closed forms could come along with a possible curvature contribution, which could either vanish due to the values of the metric fluxes, be non-zero and anisotropic, or non-zero and isotropic. In the first two cases, scale separation can be achieved parametrically, but $F_0$ would conflict with flux quantization. For isotropic curvature, scale separation would be obstructed, as argued in previous sections.
So either way, a dS solution in the parametric limit would either face a problem with flux quantization or fail to realize scale separation.

On the other hand, positive contributions at the vacuum do not arise for O5- and O6-planes, making dS solutions impossible without the inclusion of additional terms in the scalar potential, see Table \ref{tablevacuum}.
This is in agreement with no-go theorems in \cite{Andriot:2022xjh}.
Thus, we have argued that parametrically controlled dS vacua in any $3\leq D \leq 7$ cannot exist with only RR fluxes, $H_3$ flux and O-planes with a leading O$p$-plane with $2\leq p \leq 6$.

\begin{table}[ht]
\centering
\renewcommand{\arraystretch}{1.5}
\setlength{\tabcolsep}{6pt} 
\small 
    \begin{tabular}{|c|c|c|c|c|}
       \hline
       \( Op \) & \( F_{6-p} \) & other RR fluxes & EFT dimensions & Scalar Potential \\
       \hline
        \hline
        \( O2 \) & \( F_4 \) & \( \{F_0, F_2, F_6\} \)  & \( D=3 \) & \( \langle V\rangle \sim 2V_{RR}^0+V_{RR}^2-V_{RR}^{6} \) \\
        \hline
        \( O3 \) & \( F_3 \) & \( \{F_1, F_5, F_7\} \) & \( D=3 \) & \( \langle V\rangle \sim V_{RR}^1-V_{RR}^5-2V_{RR}^7 \) \\
        \( O3 \) & \( F_3 \) & \( \{F_1, F_5\} \) & \( D=4 \) & \(\langle V\rangle \sim V_{RR}^1-V_{RR}^5 \) \\
        \hline
        \( O4 \) & \( F_2 \) & \( \{F_0,F_4,F_6\} \) & \( D=3 \) & \(\langle V\rangle \sim V_{RR}^0-V_{RR}^4-2V_{RR}^6\) \\
        \( O4 \) & \( F_2 \) & \( \{F_0,F_4,F_6\} \) & \( D=4 \) & \(\langle V\rangle \sim V_{RR}^0-V_{RR}^4-2V_{RR}^6 \) \\
        \( O4 \) & \( F_2 \) & \( \{F_0,F_4\} \) & \( D=5 \) & \(\langle V\rangle \sim V_{RR}^0-V_{RR}^4 \) \\
        \hline
        \( O5 \) & \( F_1 \) & \( \{F_3, F_5, F_7\} \)  & \( D=3 \) & \( \langle V\rangle < 0\,, \text{ AdS} \) \\
        \( O5 \) & \( F_1 \) & \( \{F_3, F_5\} \)  & \( D=4 \) & \( \langle V\rangle < 0\,, \text{ AdS} \) \\
        \( O5 \) & \( F_1 \) & \( \{F_3, F_5\} \) & \( D=5 \) & \( \langle V\rangle < 0\,, \text{ AdS} \) \\
        \( O5 \) & \( F_1 \) & \( \{F_3\} \)& \( D=6 \) & \( \langle V\rangle < 0\,, \text{ AdS} \) \\
        \hline
        \( O6 \) & \( F_0 \) & \( \{F_2, F_4, F_6\} \)  & \( D=3 \) & \( \langle V\rangle < 0\,, \text{ AdS} \) \\
        \( O6 \) & \( F_0 \) & \( \{F_2, F_4, F_6\} \) & \( D=4 \) & \( \langle V\rangle < 0\,, \text{ AdS} \) \\
        \( O6 \) & \( F_0 \) & \( \{F_2, F_4\} \)  & \( D=5 \) & \( \langle V\rangle < 0\,, \text{ AdS} \) \\
        \( O6 \) & \( F_0 \) & \( \{F_2, F_4\} \)  & \( D=6 \) & \( \langle V\rangle < 0\,, \text{ AdS} \) \\
        \( O6 \) & \( F_0 \) & \( \{F_2\} \)& \( D=7 \) & \( \langle V\rangle < 0\,, \text{ AdS} \) \\
        \hline
    \end{tabular}
    \caption{This table shows, in the last column, the value of the scalar potential as calculated in equation \eqref{eq:VAdS}, before taking into account the Bianchi identity.
    We include all internal fluxes allowed by the dimension of the internal space.
    For O$p$-planes with $2 \leq p \leq 4$, it is in principle possible to realize any maximally symmetric spacetime: dS, Minkowski or AdS. In contrast, for $5 \leq p \leq 6$, solutions supported by an O$p$-plane, NSNS and RR fluxes necessarily lead to AdS solutions.}\label{tablevacuum}
\end{table}

In most of this paper, we are interested in solutions with a leading O$p$-plane that is leading in the sense that $\langle V \rangle \sim V_{Op}$. Given that the O$p$-plane charge is rather generically canceled by $H_3 \w F_{6-p}$ and this implies a similar cancellation of the potential contribution $V_{Op}$ by $V_H + V_{RR}^{6-p}$, we do need another term that is also leading. Above this could be any of the $V_{RR}^q$ with $q>6-p$. From equation \eqref{eq:VAdS} we see that the corresponding scale-separated vacua are then necessarily all AdS.

\section{\texorpdfstring{Classifying scale-separated solutions}{}}\label{sec:classificationscalesep}
In this section we initiate a full classification of all possible solutions that could potentially lead to parametrically scale-separated due to a leading-order orientifold term in the scalar potential. As we have shown above, such solutions in the weak coupling or large volume regime are automatically scale separated, as long as an unbounded flux exists.
We will be working with an approximation to the equation of motion, where we treat the orientifold as `smeared' and the fluxes as constant, see subsection 4.1 in \cite{Junghans:2023lpo} for a wonderful discussion of this approximation.
We first consider all possible cases in type IIA in subsection \ref{ssec:IIAclassification} and then type IIB in subsection \ref{ssec:IIBclassification}. We assume the absence of curvature or metric fluxes that generically prohibit scale separation due to equation \eqref{eq:VrovermKK}.
We will not consider the inclusion of NS5/ONS5 sources since it is not possible to cancel their charges with fluxes (unless they wrap trivial cycles).

Let us mention here that all existing solutions that are scale-separated and have a non-zero cosmological constant have either no supersymmetry or preserve at most four supercharges. It would be interesting to understand the reason for this and try to prove the absence of scale-separated solutions in the presence of larger amounts of supersymmetry along the lines of \cite{Cribiori:2023ihv, Cribiori:2022trc, Montero:2022ghl, Perlmutter:2024noo}. Our analysis is agnostic about the amount of preserved or broken supersymmetry and applies generically to any geometric compactification. We assume we have a leading O$p$-plane whose contribution to the scalar potential is of the order of the inverse length scale squared and non-vanishing $V_{Op} \sim 1/L_{\langle V \rangle}^2 \neq 0$. In the large volume limit we are interested in, this prevents the presence of a $V_{Op'}$ term from O$p'$ planes with $p'>p$ since the latter would dominate over the lower dimensional O$p$-plane, see equation \eqref{eq:VOpovermKK}. 

\subsection{Limits of our analysis}

In our analysis, we focus on the universal features of $D$-dimensional effective theories and examine their behavior using the equations of motion for the dilaton and the overall volume. We do not delve into the details of the internal space, which is very model dependent. Therefore, our analysis captures the universal features of any setup by studying the universal equations of motion that have to hold for any geometric compactification. However, this approach provides necessary but not always sufficient conditions, as a detailed compactification involves additional equations of motion, related to fluctuations of internal space components, which may not always be satisfied by our flux ansatz for a specified internal geometry. Furthermore, explicit geometric spaces might not allow for all the fluxes that are needed to cancel the O-plane charge and give scale separation because there are no appropriate cycles that are even or odd under the orientifold projection. Lastly, for the dS vacua we discuss below, one would also have to ensure the stabilization of all other geometric moduli to avoid a runaway resulting from subleading corrections.

Nonetheless, as we have previously demonstrated and will further explore using the scaling analysis below, the universal study of type II compactifications with a leading orientifold and specific types of fluxes accurately captures the vacuum and parametric behavior of more detailed constructions such as the DGKT setup \cite{DeWolfe:2005uu} and G2 compactifications of massive type IIA \cite{Farakos:2020phe}. Building on this, we present new insights into the possibility of scale-separated solutions within the universal approach, which merit further investigation in more detailed setups. In particular, we demonstrate that scale-separated solutions with O6-planes and parametric control across various dimensions may be feasible, though their realization is likely to require more intricate internal geometries, if such constructions are possible at all.
We also observe parametric control and scale separation in setups involving O3-, O4-, and O5-planes,\footnote{Recall that for O3-planes the 10d string coupling $e^\phi$ cannot be made parametrically small as discussed in subsection \ref{ssec:weaklargelimit}. So, in that case only parametric control refers to a large volume and a large $D$ dimensional dilaton $\tau$ only.} achieved using only harmonic fluxes, which may offer valuable hints toward constructing more explicit examples.

\subsection{\texorpdfstring{Classifying scale-separated solutions in type IIA}{}}\label{ssec:IIAclassification}
In this subsection we discuss compactifications of type IIA to $D$ spacetime dimensions. We want parametrically scale-separated solution with a leading scalar potential term from an O2, O4, or O6-plane.\footnote{Recall that we are canceling the O$p$-planes charge using the $H_3 \w F_{6-p}$ term in equation \eqref{Bianchi1}. This requires that $p\leq 6$.} We require the $D$ dimensional spacetime to be maximally symmetric so that the sources all have to be spacetime filling, which means for an O$p$-plane that $p+1\geq D$.

\subsubsection{\texorpdfstring{O2-planes in effective theories in \boldmath{$D=3$}}{}}\label{sssec:O2}

Since we are considering spacetime-filling O-planes, we begin with O2-planes, which can be spacetime-filling only in $D=3$ effective theories. 
However, as argued in Section~\ref{ssec:weaklargelimit}, parametrically classical solutions cannot be achieved in setups with leading O2-planes.

The minimization of a setup with leading O2-planes and all possible fluxes can potentially lead to Minkowski, AdS, or dS solutions, as shown in Table \ref{tablevacuum}.
We always include the fluxes $H_3$ and $F_4$, which are essential to cancel the tadpole. The inclusion of $F_0$ and $F_2$ can potentially lead to dS solutions. However, as explained in subsection~\ref{ssec:nodS}, when leading, these quantized fluxes prevent the existence of parametric solutions. It should be noted that even if one were interested in solutions without parametric behavior, the inclusion of the Romans mass $F_0$ is not possible, as this would require $H_3 = 0$ in order to satisfy the Bianchi identity $0 = H_3 \wedge F_0$ in equation \eqref{Bianchi2}. Thus, for parametrically scale-separated solutions $F_0=0$ and $F_2$ is zero or subleading and we can neglect it. Then, the only remaining leading flux is $F_6$, which leads to AdS solutions, and the parametric limit can be taken without constraints from flux quantization.

Since, in our construction, the fluxes are considered harmonic, $F_6$ appears only in the flux equations of motion in \eqref{flux2} as $0 = \star F_6 \wedge F_4$ and must remain unbounded using one of the methods we proposed in Section \ref{sec:unboundfluxes}. The vacuum expectation value of the scalar potential for these vauca is given by
\begin{equation}
    \langle V\rangle=-\frac{1}{4}V^6_{RR} \,.
\end{equation}
Performing the scaling analysis to study the parametric behavior of this potential construction, we find that while we can parametrically go to large volume, we simultaneously approach strong string coupling  
\begin{align}\label{propO2}
    \rho\sim N^r\sim N^{f_6}\,,\quad\quad
    e^{\phi}\sim N^{\frac{1}{2}f_6} \,,\quad\quad
    \tau\sim N^{\frac{5}{2}f_6} \,,\quad\quad
    \frac{L^2_{KK}}{L^2_{AdS}}\sim N^{-2f_6} \,,
\end{align}  
which would then require an uplift to M-theory. This is consistent and expected from our general analysis in subsection \ref{ssec:weaklargelimit}.

Let us point out that, if we had also considered a leading $F_2$ flux, which could give dS vacua, its scaling would be $F_2\sim N^{-f_6}$. So, we see explicitly that fluxes that could lead to parametric classical dS solutions cannot be quantized in the parametric regime, confirming the general analysis of subsection~\ref{ssec:nodS}.

It is also clear that, with $F_6$ turned off, the setup including only the fluxes required for tadpole cancellation leads to a Minkowski vacuum, as discussed in subsection \ref{ssec:nodS}. This was studied in detail in \cite{Farakos:2020phe}, where the internal space was chosen to be a toroidal orbifold with G2 structure. It was shown that the resulting full solution, including the G2 metric deformations, and not just the universal volume modulus, leads to a Minkowski vacuum when O2-planes are dominant. In such solutions, the scalar potential vanishes and scale separation arises automatically. However, these are not the type of solutions we are interested in this paper, as we focus on cases where a leading O-plane contribution gives rise to a non-zero scalar potential. The paper \cite{Farakos:2020phe} also discusses the inclusion of O6-planes, which modify equation \eqref{Bianchi2} by introducing an O6-plane source term. However, the O6-plane would dominate over the O2-plane in the large-volume limit. This is discussed below in subsection \ref{sssec:O6}.

\subsubsection{\texorpdfstring{O4-planes in effective theories in \boldmath{$D=3,4,5$}}{}}\label{sssec:O4}

For constructions involving O4-planes that make a leading contribution to the scalar potential, we recall that the spatial involution $\sigma$ that is part of the orientifold projection acts on the mass parameter in type IIA via $\sigma(F_0) = -F_0$. Thus, for constant fluxes, this requires us to set $F_0=0$ and according to Table \ref{tablevacuum}, allowing for either $F_4$, $F_6$, or both can lead to AdS solution
\begin{align}
    \langle V\rangle=-\frac{D-2}{D+3}\left(V_{RR}^4+2V_{RR}^6 \right) \,,
\end{align}
where we have already incorporated the $F_2$ and $H_3$ fluxes that cancel the tadpole. 

The Bianchi identities in equation \eqref{Bianchi1} and the flux equations of motion which need to be satisfied, see equations \eqref{flux1} and \eqref{flux2}, take the form
\begin{align}
    0&=H_3\wedge F_2+\mu_{O4}J_5\,, \label{df1O4}\\
    0&=H_3\wedge F_4 \,, \\ 
    0&=H_3\wedge \star F_{4} \,,\label{FO41} \\
    0&=H_3\wedge \star F_{6} \,, \label{FO42} \\
    0&=F_2\wedge \star F_{4}+F_4\wedge \star F_{6} \,. \label{FO43}
\end{align}
Above we have used that $H_3 \w F_6$ vanishes automatically for internal $H_3$ and $F_6$ fluxes in an internal $10-D$ space with $D=3,4,5$. 

In $D=3$, both $F_4$ and $F_6$ can be expanded in harmonic forms of the internal space while respecting the parity conditions imposed by the orientifold projection $\sigma(F_4) = -F_4$ and $\sigma(F_6) = +F_6$. 
For $D=4$, $F_6$ gets projected out if the volume form of the internal space is odd under the O4 orientifold projection. 
For $D=5$, we cannot have an internal $F_6$ flux and we are left again with only $F_4$ fluxes, provided the geometry supports odd 4-cycles that can be threaded with fluxes.

\paragraph{Scaling analysis}
Now we perform a scaling analysis using \eqref{Rscaling}-\eqref{NSlanescaling}, considering that the fluxes related to the tadpole are leading and thus, $f_2=-h_3=0$, we find that the following relations for the scaling should hold: The fluxes scale in the following way
\begin{align}
    F_4\sim N^{f_4}
    \,,\quad\quad 
    F_6\sim N^{f_6}\sim N^{2f_4} \,,
\end{align}
while the volume, string coupling and ratio of length scales behave as
\begin{align}\label{propO4}
    \rho\sim N^r\sim N^{f_4}\,,\quad\quad
    e^{\phi}\sim N^{-\frac{1}{2}f_4}
    \,,\quad\quad 
    \tau\sim N^{\frac{1}{2}\frac{12-D}{D-2}f_4}
    \,,\quad\quad 
    \frac{L^2_{KK}}{L^2_{AdS}}\sim N^{-2f_4} \,.
\end{align}
Note that the scaling for the quantities above (except $\tau$) are independent of the spacetime dimension $D$. 
If one of the fluxes, either $F_4$ or $F_6$, is unbounded, the solutions can be at parametrically large volume, weak coupling, and they are parametrically scale-separated as expected from our general analysis.  
In cases where $F_4$ is subdominant or zero but we have an unbound $F_6$ flux, we simply have to replace $f_4$ with $\tfrac12f_6$ in the formulas in equation~\eqref{propO4}.

\paragraph{Stability} To assess the stability of these general solutions, we check when the eigenvalues $\lambda_{+}, \lambda_{-}$ of the Hessian in \eqref{Hessian} are positive\footnote{In general AdS vacua are stable if the masses squared are above the Breitenlohner-Freedman bound \cite{Breitenlohner:1982bm, Breitenlohner:1982jf}. However, we list here the requirement for them to be positive since the equations are simpler.} in generic dimension $D$ for a scalar potential incorporating O4-planes and all relevant fluxes. In the following, we list conditions that ensure that both eigenvalues are positive, while considering the maximum number of quantized RR fluxes that can be taken to be large, $F_4$ and $F_6$. We find two main conditions, and either one of them should be satisfied. The first one is
\begin{align}
    V_{RR}^4\leq 2(D+1)V_{RR}^6 \,.
\end{align}
If this is not satisfied and $V_{RR}^4> 2(D+1)V_{RR}^6\,$, then we need also that
\begin{align}
    V_H> \frac{D}{3+D}\frac{(V_{RR}^4+2V_{RR}^6)(V_{RR}^4-2(D+1)V_{RR}^6)}{(2D+1)V_{RR}^4+2(3D-1)V_{RR}^6} \,.
\end{align}
For $F_4$ subleading the first condition becomes trivial. If $F_6$ is subleading the second condition simplifies but remains non-trivial.

\subsubsection{\texorpdfstring{O6-planes in effective theories in \boldmath{$D=3,4,5$} }{}}\label{sssec:O6}

Next, we examine $D$-dimensional theories with an O6-plane and constrain the possible fluxes that can be included, in order to obtain AdS vacua with the desired characteristics.

For $D = 7$, the only fluxes compatible with the dimensions of the internal space are those related to tadpole cancellation, since the only RR flux that can be expanded in harmonic forms of the internal space is $F_2$, which must be set to zero in order to satisfy the equation $\star F_2 \wedge F_0 = 0$. As a result, the only solution we can obtain is a Minkowski vacuum.
For $D=6$, the $F_4$ flux needs to be proportional to the internal volume form. The equation $0=H_3 \wedge\star F_4$ can then only be satisfied by setting $F_4$ to zero or by including O2-planes. However, O2-planes would break the maximal symmetry of the $D=6$ dimensional external space. Similarly, the equation $0 = \star F_2 \wedge F_0$ requires $F_2$ to be zero. As a result, we are left only with the fluxes involved in the tadpole cancellation, and the resulting vacuum will be a Minkowski solution.
For $D=5$, the internal fluxes compatible with the dimensions of the internal space are $H_3$, $F_0$, $F_2$ and $F_4$. We can in principle get parametrically controlled AdS solution from parametrically large $F_2$ and/or $F_4$ fluxes. A potential explicit 5D construction would require an internal space admitting 1-, 3-, 4-, and 5-forms.

For $D=3,4$, it is known that smeared, classical, and scale-separated solutions with full moduli stabilization exist, as shown in \cite{DeWolfe:2005uu, Farakos:2020phe}. In the more general compactifications considered in these references, where all type IIA fluxes or combinations thereof could potentially be accommodated, one must satisfy the full set of equations of motion \eqref{Bianchi1}–\eqref{flux2}, making the analysis significantly more involved. We write down the general expression for the vacuum expectation value
\begin{equation}
    \langle V\rangle=-\frac{D-2}{D+5}\left(V_{RR}^2+2V_{RR}^4+3V_{RR}^6\right) \,,
\end{equation}
which agrees with the vacuum expectation values in \cite{Farakos:2020phe,DeWolfe:2005uu} for $D=3,4$.

\paragraph{Scaling analysis}
According to what we discussed above, the following analysis is valid for $D = 3, 4, 5$, where for $D = 5$, one should set $F_6 = 0$.
We perform a scaling analysis using \eqref{Rscaling}-\eqref{NSlanescaling}, considering that the fluxes related to the tadpole are leading and thus, $f_0=h_3=0$.
The fluxes scale in the following way
\begin{align}
    F_2\sim N^{f_2}\sim N^{\frac{1}{2}f_4}
    \,,\quad\quad 
    F_4\sim N^{f_4}
    \,,\quad\quad 
    F_6\sim N^{f_6}\sim N^{\frac{3}{2}f_4} \,,
\end{align}
while the volume, string coupling and ratio of length scales behave as
\begin{align}\label{propO6}
    \rho\sim N^r\sim N^{\frac{1}{2}f_4}\,,\quad\quad
    e^{\phi}\sim N^{-\frac{3}{4}f_4}
    \,,\quad\quad
    \tau\sim N^{\frac{1}{4}\frac{16-D}{D-2}f_4}
    \,,\quad\quad
    \frac{L^2_{KK}}{L^2_{AdS}}\sim N^{-f_4} \,.
\end{align}
If any of the $F_2$, $F_4$ or $F_6$ fluxes are unbounded, the solutions contain regions of parametrically large volume, weak coupling, and scale separation. 

\paragraph{Stability} To assess the stability of these general solutions, we compute the eigenvalues $\lambda_{+}, \lambda_{-} > 0$ of the Hessian in \eqref{Hessian} in generic dimension $D$ for the scalar potential incorporating O6-planes and all relevant fluxes. In the following, we list the conditions that ensure that both eigenvalues remain positive while considering all possible fluxes. We identify two main conditions, and one of them should be satisfied. The first one is
\begin{align}
    V_{RR}^2\leq \frac{2}{3}(D-1)V_{RR}^4+(2D+1)V_{RR}^6 \,.
\end{align}
If this is violated and $V_{RR}^2> \frac{2}{3}(D-1)V_{RR}^4+(2D+1)V_{RR}^6$ then we need
\begin{align}
    V_H> \frac{D}{5+D}\frac{\big(V_{RR}^2+2V_{RR}^4+3V_{RR}^6\big)\big(3V_{RR}^2+2V_{RR}^4-3V_{RR}^6-2D(V_{RR}^4+3V_{RR}^6)\big)}{3V_{RR}^2+2V_{RR}^4-3V_{RR}^6+2D(V_{RR}^2+3V_{RR}^4+6V_{RR}^6)} \,.
\end{align}
For $F_2$ subleading, the first condition becomes trivial and $\rho$ and $\tau$ automatically have positive masses squared. However, if either $F_4$ or $F_6$ are subleading, then there remains a non-trivial condition.

\subsection{\texorpdfstring{Classifying scale-separated solutions in type IIB}{}}\label{ssec:IIBclassification}

In this subsection, we study AdS solutions from type IIB with spacetime-filling O-plane sources. We find for O5-planes that the universal moduli can be stabilized, while the string coupling and volume can be made parametrically small and large, respectively. This is due to unbounded fluxes which can be either $F_3$ or $F_5$. We also find that for space-filling O3-planes, there exist parametric AdS solutions. However, as argued in Section~\ref{ssec:weaklargelimit}, the ten-dimensional string coupling does not scale at all but could potentially be numerically small.
The solutions have parametrically large volume, and as argued in equation \eqref{eq:VOpovermKK}, this ensures parametric scale separation.

\subsubsection{\texorpdfstring{O3-planes and \boldmath{$D=3,4$} effective theories}{}}

We want the O3-planes to be always spacetime filling, this means that since their worldvolume is 4-dimensional, the dimensions of the external space should always be $3\leq D\leq 4$. 
We neglect the $F_1$ flux, since it cannot be leading for parametric solutions as discussed in subsection \ref{ssec:NoquantizedRRforsmallq}.
The internal flux $F_7$ is proportional to the internal volume form for $D = 3$ and therefore has to vanish due to equation \eqref{flux1} that requires $H_3 \w \star F_7=0$. For $D = 4$, the $F_7$ flux cannot fit in the internal space, so we exclude $F_7$ from our constructions.
Thus, to obtain AdS solutions, we can rely solely on the $F_5$ flux and we write down the vacuum expectation value
\begin{align}\label{vevO3}
    \langle V\rangle=-\frac{D-2}{D+2}V_{RR}^5 \,,
\end{align}
and the non-trivial equations to satisfy are
\begin{align}
    0&=H_3\wedge\star F_{3} + \mu_{O3}J_6 \,, \\
    0&=H_3\wedge\star F_{3} \,, \\
    0&=H_3\wedge\star F_{5} \,, \\
    0&=F_3\wedge\star F_{5} \,.
\end{align}
\paragraph{Scaling analysis}
Now we perform a scaling analysis using \eqref{Rscaling}-\eqref{NSlanescaling}, considering that the fluxes related to the tadpole are leading and thus, $f_3=h_3=0$.
We find that the following relations for the scaling behavior should hold
\begin{align}
    F_5\sim N^{f_5} \,,
\end{align}
while the volume, string coupling and ratio of length scales behave as
\begin{align}\label{propO3}
    \rho\sim N^r\sim N^{f_5}
    \,,\quad\quad
    e^{\phi}\sim N^{0}
    \,,\quad\quad 
    \tau\sim N^{\frac{1}{2}\frac{D-10}{2-D}f_5}
    \,,\quad\quad 
    \frac{L^2_{KK}}{L^2_{AdS}}\sim N^{-2 f_5} \,.
\end{align}
The 10d string coupling is independent of the flux content and does not vary parametrically, whereas the other quantities exhibit parametric behavior, large volume and scale separation.
This aligns perfectly with our general analysis in \eqref{eq:scaleseparation} that scale separation can be achieved as long as the system is at large volume or weak coupling. Although in these solutions the string coupling does not scale parametrically, this does not mean it cannot be numerically small due to details of the compactification.

\paragraph{Stability}
We consider O3-planes for the setup we discussed above and include $F_3$, $F_5$ and $H_3$ fluxes.
We find the squared masses of the two scalar fields from equation \eqref{eq:massessquared} to be
\begin{align}
\lambda_{+} = \frac{2D}{10-D} V_{RR}^5\,,
\quad
\lambda_{-} = 2 V_H\,,
\end{align}
both of which are always positive without any constraints.

\subsection{\texorpdfstring{O5-planes and \boldmath{$D=3,4,5,6$} effective theories}{}}

We require the O5-planes to be spacetime-filling, which implies that their six-dimensional worldvolume must include the entire external spacetime. As a result, the allowed dimensions for the external spacetime are $3 \leq D \leq 6$.  

To satisfy the tadpole cancellation condition, we include $F_1$ flux and $H_3$ fluxes. For $D = 3$, $F_3$ and $F_5$ fluxes can be included if the internal space contains the relevant forms that are even and odd under the orientifold projection, respectively. However, $F_7$ would have to be proportional to the internal volume form and the flux equation~\eqref{flux1} then becomes $0 = H_3 \wedge \star F_7$, which forces $F_7$ to be zero. It is obvious that for $D=4, 5, 6$ the $F_7$ flux cannot fit into the internal space and therefore we do not consider $F_7$ at all.

There are recent attempts to construct scale-separated $D=3$ vacua from type IIB with O5-planes compactified on spaces that are not Ricci flat \cite{Emelin:2021gzx, Arboleya:2024vnp, VanHemelryck:2025qok}. In these setups it was not possible to incorporate $H_3$ flux, and the tadpole cancellation was achieved by using metric fluxes. In our constructions we present a potentially different approach\footnote{Recall that metric fluxes and $H_3$ fluxes can be related by T-duality. So, the approaches might be T-dual and therefore the same.} where the fluxes are harmonic and $H_3$ must be non-zero to cancel the tadpole.  We remain agnostic about the internal space that could allow for this. 

Thus, in our setup, we are left with the fluxes that cancel the tadpole, along with the additional RR fluxes $F_3$ and $F_5$, which can potentially be incorporated in $D=3,4,5$ constructions. This means the resulting vacuum expectation value of the scalar potential is
\begin{align}\label{vevO5}
    \langle V\rangle = -\frac{D-2}{D+4}\left(V_{RR}^3 + 2V_{RR}^5\right) \,.
\end{align}  
and the non-trivial remaining equations are
\begin{align}
    0&=H_3\wedge F_{1} + \mu_{5} J_{4} \,,\\
    0&=H_3\wedge \star F_{3} \,,\\
    0&=H_3\wedge \star F_{5} \,,\\
    0&=\star F_3\wedge F_{1}+\star F_5\wedge F_{3} \,.
\end{align}
For $D=6$ the $F_5$ flux necessarily vanishes and we have only $F_3$ that can be potentially made parametrically large.

\paragraph{Scaling analysis}
Now we perform a scaling analysis using \eqref{Rscaling}-\eqref{NSlanescaling}, considering that the fluxes related to the tadpole are leading and thus, $f_1=h_3=0$, we find that the following relations for the scaling behavior should hold
\begin{align}
    F_3\sim N^{f_3}\sim N^{\frac{1}{2}f_5}
    \,,\quad\quad
    F_5\sim N^{f_5}\sim N^{2f_3} \,,
\end{align}
while the volume, string coupling and ratio of length scales behave as
\begin{align}\label{propO5}
    \rho\sim N^r\sim N^{\frac{1}{2}f_5}\,,\quad\quad
    e^{\phi}\sim N^{-\frac{1}{2}f_5}
    \,,\quad\quad 
    \tau\sim N^{\frac{1}{4}\frac{14-D}{D-2}f_5}
    \,,\quad\quad 
    \frac{L^2_{KK}}{L^2_{AdS}}\sim N^{-f_5} \,.
\end{align}
If the $F_5$ flux is absent or subdominant, then we have to change in the equations above $f_5 \to 2 f_3$.

\paragraph{Stability}
Below, we list the conditions that ensure that both eigenvalues are positive. The first condition is
\begin{align}
    V_{RR}^3\leq D V_{RR}^5 \,.
\end{align}
If this condition is violated and $V_{RR}^3> DV_{RR}^5$, then we have to impose
\begin{align}
       V_H>\frac{D}{D+4}\frac{(V_{RR}^3+2V_{RR}^5)(V_{RR}^3-DV_{RR}^5)}{V_{RR}^3(D+1)+3DV_{RR}^5} \,.
\end{align}
Again, it is useful to note that, when both leading $F_3$ and $F_5$ are present, there is always a non-trivial condition to satisfy. However, for $3\leq D\leq 5$, for $F_3$ subleading or zero, the first condition becomes trivial and the solutions are guaranteed to be non-tachyonic.

\section{Conclusion}\label{sec:Conclusion}

In this paper, we studied $D$-dimensional effective theories arising from type II compactifications with O-planes. Using the equations of motion and scaling analysis, we examined the universal properties of (A)dS solutions and explored the possibility of parametric control and scale separation.
We investigated the interplay between scale separation and the presence of leading ingredients in the scalar potential such as curvature, fluxes, and local sources.
We found that the presence of leading O-planes, which is crucial for constructing such solutions, points to parametric regimes characterized by weak coupling, large volume, or both, necessary for achieving parametric scale separation.
On the other hand, by imposing tadpole cancellation through fluxes, we derived constraints on parametric control for leading O2- and O3-planes, while no constraints were found for O4-, O5-, and O6-planes.
We also showed that non-vanishing internal curvature from isotropic spaces obstructs scale separation, and we related this to the generic absence of gap in the spectrum of CFTs that arise in the AdS/CFT context. Furthermore, we highlighted the conflict between parametric control and flux quantization, particularly in the context of fluxes associated with potential dS solutions.
We examined the parametric behavior and stability of universal AdS solutions arising from compactification and classified them according to the dimensionality of the O-planes. Through this universal analysis, we concluded that there are potentially constructions that could lead to parametrically classical solutions with scale separation across several dimensions. The explicit construction of such solution is non-trivial and a more detailed study of such solutions might even produce obstructions. However, our results summarized in Table \ref{tablevacuum} pave the way to a full classification of (A)dS solutions from flux isotropic flux compactifications in the absence of curvature.

As a future direction, it would be interesting to formulate more general statements in more complex settings. For example, we could investigate whether anything can be said about the presence of metric fluxes, non-harmonic forms, and anisotropic spaces, and consider including other ingredients such as Casimir effects or KK monopoles, in order to make our conclusions more general.
It would also be essential, though non-trivial, to explore more explicit examples by attempting to compactify over detailed internal geometries, with characteristics guided by the hints for the potential existence of such solutions provided by our general analysis.
Given the indications of potentially more scale-separated solutions than those currently known, it would be particularly interesting to understand such solutions from the AdS/CFT perspective.

\section*{Acknowledgments}
We would like to thank Muthusamy Rajaguru for his initial collaboration, and David Andriot, Fotis Farakos, Daniel Junghans, Vincent Van Hemelryck, and Thomas Van Riet for useful discussions and comments on the draft.
This work is supported in part by the NSF grant PHY-2210271 and the Lehigh University CORE grant with grant ID COREAWD40.

\appendix
\section{Review of type II supergravity}\label{app:typeII}

The bosonic part of the type II supergravity action, the NSNS and RR parts, have the following form in string frame
\begin{align}\label{stringaction}
    S&=\frac{1}{2\kappa_{10}^2}\int \text{d}^{10}X\sqrt{-G_S}e^{-2\phi}\left(R_{10}+4(\partial\phi)^2
    - \frac{1}{2}\vert H_3\vert^2
    - \frac{1}{4}e^{2\phi}\sum_{q}\vert F_q\vert^2
    \right)\,,
\end{align}
where $2\kappa^2_{10}=(2\pi)^7\alpha^{\prime 4}$ with $\alpha^{\prime}=l_s^2$. The form degrees $q$ of the RR field strengths $F_q$ depend on the type II theory and the formalism, see equation \eqref{pq}. 
For the O$p$-planes, the DBI and Wess-Zumino part, are given by
\begin{align}\label{sources}
    S_{loc}&=-T_p\int\text{d}^{p+1}X e^{-\phi}\sqrt{-P[G_S]}+T_{p}\int C_{p+1}\,,
\end{align}
where $P[G_S]$ is the determinant of the pull-back of the ten-dimensional metric $G_{S,MN}$.
For our purposes, we will need the equations of motion in the 10d Einstein frame, so we perform the following Weyl rescaling 
\begin{equation}
    G_{S,MN}=e^{\phi/2}G_{MN} \,,
\end{equation}
and the bosonic action together with the DBI part of local objects takes the form
\begin{equation}
\begin{split}
    S_{E}&=\frac{1}{2\kappa_{10}^2}\int \text{d}^{10}X\sqrt{-G}\left(R_{10}-\frac{1}{2}(\partial\phi)^2
    - \frac{1}{2}e^{-\phi}\vert H_3\vert^2
    - \frac{1}{4}e^{\frac{5-q}{2}\phi}\sum_{q}\vert F_q\vert^2
    \right)\, \\
    &-T_p\int\text{d}^{10}X \sqrt{-G}\,e^{\frac{p-3}{4}\phi}\delta(\Sigma_i)
    \,.
\end{split}
\end{equation}
We also use the following index notation: $M, N = 1, \dots, 10$, while Greek letters $\mu, \nu$ label the coordinates of the $D$-dimensional external space, and lowercase Latin letters $m, n$ denote the indices of the $10 - D$ internal space.
In most of our analysis, we will consider Ricci-flat spaces, where the Ricci tensor of the internal space vanishes. However, since we also wish to comment on internal spaces with group structure, we will, for now, take the Ricci scalar to be  
\begin{equation}  
    R_{10} = R_{D} + R_{(10-D)} + \dots \,,  
\end{equation}  
where the dots represent derivatives of the warp factors, which vanish under the smearing approximation.

The variation with respect to the ten-dimensional metric, $\delta S_E/\delta G^{MN}=0$, gives the Einstein equation
\begin{equation}\label{Einstein1}
\begin{split}
    &\left(R_{MN}-\frac{1}{2}G_{MN}R_{10}\right)
    -\frac{1}{2}\left(\partial_M\phi\partial_N\phi-\frac{1}{2}G_{MN}(\partial\phi)^2\right) \\
    &-\frac{1}{2}e^{-\phi}\left(\vert H_3\vert^2_{MN}-\frac{1}{2}G_{MN}\vert H_3\vert^2\right)
    -\frac{1}{4}e^{\frac{5-q}{2}\phi}\sum_{q}\left(\vert F_q\vert^2_{MN}-\frac{1}{2}G_{MN}\vert F_q\vert^2\right)\\
    &
    +\frac{1}{2}e^{\frac{p-3}{4}\phi}T^{loc}_{MN}=0 \,,
\end{split}
\end{equation}
where we have defined the stress-energy tensor of the O-planes as
\begin{align}\label{sourcestress}
    T_{MN}=2\kappa_{10}^2T_pP[G]_{MN}\delta_{\Sigma_i} \,,
\end{align}
and have extracted any dilaton arising from the Weyl rescaling, making it explicit in the equations of motion.
For later convenience, we find it useful to write down the following relations
\begin{align}\label{sourcestress2}
    T^{loc}
    =G^{MN}T^{loc}_{MN}
    &=2\kappa^2_{10}T_p(p+1)\,\delta_{\Sigma_i}\,, \\
    G^{\mu\nu}T^{loc}_{\mu\nu}
    &=2\kappa^2_{10}T_p\,D\,\delta_{\Sigma_i}\,, \\
    G^{mn}T^{loc}_{mn}
    &=2\kappa^2_{10}T_p(p+1-D)\,\delta_{\Sigma_i} \,.
\end{align}
Now we contract the Einstein equation in \eqref{Einstein1} with the ten-dimensional metric and obtain
\begin{align}\label{Ricci10}
    &R_{10}
    =
    \frac{1}{2}(\partial\phi)^2
    +\frac{1}{4}e^{-\phi}\vert H_3\vert^2
    +\frac{5-q}{16}\sum_qe^{\frac{5-q}{2}\phi}\vert F_q\vert^2
    +\frac{1}{8}e^{\frac{p-3}{4}\phi}T^{loc}\,,
\end{align}
and then we plug \eqref{Ricci10} back into \eqref{Einstein1} and obtain the 10-dimensional trace-reversed Einstein equation
\begin{align}\label{EinsteinTR}
    &R_{MN}
    -\frac{1}{2}\partial_M\phi\partial_N\phi
    -\frac{1}{2}e^{-\phi}\left(\vert H_3\vert^2_{MN}
    -\frac{1}{4}G_{MN}\vert H_3\vert^2\right) \\
    &-\frac{1}{4}\sum_qe^{\frac{5-q}{2}\phi}\left(\vert F_q\vert^2_{MN}-\frac{q-1}{8}G_{MN}\vert F_q\vert^2\right)
    +\frac{1}{2}e^{\frac{p-3}{4}\phi}\left(T^{loc}_{MN}-\frac{1}{8}G_{MN}T^{loc}\right)=0 \,. \nonumber
\end{align}
Next, we contract this equation with the inverse $D$-dimensional external space metric $g^{\mu\nu}$ to obtain
\begin{align}\label{EinsteinD}
    R_{D}
    =
    -\frac{D}{8}e^{-\phi}\vert H_3\vert^2
    -\sum_{p}
    \frac{D(q-1)}{32}e^{\frac{5-q}{2}\phi}\vert F_q\vert^2
    -\kappa_{10}^2T_p\,D\left(\frac{7-p}{8}\right)e^{\frac{p-3}{4}\phi}\delta_{\Sigma_i} \,.
\end{align}
Taking the trace-reversed form of \eqref{EinsteinTR} and keeping only the internal space indices, we obtain
\begin{equation}\label{EinsteinT}
\begin{split}
    &R_{mn}
    -\frac{1}{2}\partial_m\phi\partial_n\phi
    -\frac{1}{2}e^{-\phi}\left(\vert H_3\vert^2_{mn}
    -\frac{1}{4}G_{mn}\vert H_3\vert^2\right) \\
    &-\frac{1}{4}\sum_{q}e^{\frac{5-q}{2}\phi}\left(\vert F_q\vert^2_{mn}-\frac{q-1}{8}G_{mn}\vert F_q\vert^2\right)
    +\frac{1}{2}e^{\frac{p-3}{4}\phi}\left(T^{loc}_{mn}-\frac{1}{8}G_{mn}T^{loc}\right)=0 \,.
\end{split}
\end{equation}
We contract this equation with the internal space metric to get
\begin{equation}
\begin{split}
    R_{(10-D)}
    &=
    \frac{1}{2}(\partial_m\phi)^2
    +\frac{2+D}{8}e^{-\phi}\vert H_3\vert^2
    +\sum_q\frac{8q-(10-D)(q-1)}{32}e^{\frac{5-q}{2}\phi}\vert F_q\vert^2\\
    &+\kappa^2_{10}T_p\left(\frac{8D-(D-2)(p+1)}{8}\right)e^{\frac{p-3}{4}\phi}\,\delta_{\Sigma_i} \,.
\end{split}
\end{equation}
Turning the RR fluxes into internal ones requires multiplying the RR terms by a factor of 2 in all the equations where they appear.

\section{O-planes and scale separation}\label{app:Oplenscaleseparation}

The presence of O-planes and their relation to scale separation have been discussed previously in \cite{Gautason:2015tig}. In that paper, the 10d equations of motion were used for the analysis, including the variation of the dilaton, while the scale separation condition was defined through the integrated ratio of the internal and external curvatures, which were determined through the equations of motion.
Here we keep the dimensions arbitrary 
and we express the internal $R_D$ and external $R_{(10-D)}$ curvatures in terms of the fluxes and sources through the equations of motion, leading to the following expression
\begin{equation}\label{ratiocurv}
    \left\vert\frac{\int R_D}{\int R_{(10-D)}}\right\vert= \frac{\frac{D}{8}k
    +\sum_{p}
    \frac{D(q-1)}{16}f_q
    +T_p\,D\left(\frac{7-p}{8}\right)s_p}{\frac{1}{2}\sigma
    +\frac{2+D}{8}k
    +\sum_q\frac{8q-(10-D)(q-1)}{16}f_q
    +T_p\left(\frac{8D-(D-2)(p+1)}{8}\right)s_p} \,,
\end{equation}
where $D$ and $10-D$ are the dimensions of the external and internal space, respectively, while
\begin{align}
    k=\int e^{-\phi}\vert H_3\vert^2\,,
    \quad
    f_q=\int e^{\frac{5-q}{2}\phi}\vert F_q\vert^2\,,
    \quad
    s_p=\kappa_{10}^2\int e^{\frac{p-3}{4}\phi}\delta_{\Sigma_i}\,,
    \quad
    \sigma=\int(\partial_m\phi)^2 \,.
\end{align}
The ratio in \eqref{ratiocurv} matches \cite{Gautason:2015tig} for $D=4$, up to a factor of $1/2$ in the sources, which depends on conventions.  
For D-branes, where $T_p > 0$, all terms in the numerator and denominator have the same sign, meaning that large flux values can only bound the ratio by a small number. As argued in \cite{Gautason:2015tig}, the only way to achieve scale separation in this case is through large dilaton variations, $\sigma \gg \textit{flux + D-brane contributions}$, which can lead to a hierarchy of scales since the ratio becomes small. However, for O-planes, where $T_p < 0$, a varying dilaton is not even necessary, as there are additional ways to make the ratio small due to the negative contribution of the O-planes. This provides a rationale for how the presence of O-planes can help achieve scale separation. It is worth noting that other sources of negative contributions, such as Casimir energy \cite{Luca:2022inb, Parameswaran:2024mrc}, could play a similar role. However, such constructions have not been extensively studied.

\bibliographystyle{JHEP}
\bibliography{refs}

\end{document}